\documentclass[12pt]{article}
\usepackage{epsf}
\usepackage{epsfig}

\def\beq{\begin{eqnarray}}
\def\eeq{\end{eqnarray}}

\begin{document}

\begin{titlepage}
\begin{flushright}
 PITHA 01/05\\
 hep-ph/0106067\\
 June 5, 2001
\end{flushright}
\vskip 2.4cm

\begin{center}
\boldmath
{\Large\bf Systematic approach to exclusive \\[0.3cm] 
$B\to V\ell^+\ell^-$, $V\gamma$ decays}
\unboldmath 
\vskip 2.2cm
{\sc M. Beneke}, \hspace*{0.3cm}{\sc Th.~Feldmann}
\hspace*{0.3cm}and\hspace*{0.3cm}{\sc D.~Seidel}
\vskip .5cm
{\it Institut f\"ur Theoretische Physik~E, RWTH Aachen, 
52056 Aachen, Germany}

\vskip 2.5cm

\end{center}

\begin{abstract}
\noindent 
We show --  by explicit computation of first-order corrections -- 
that the QCD factorization approach previously applied to hadronic 
two-body decays and to form factor ratios also allows us to compute 
non-factorizable corrections to exclusive, radiative $B$ meson 
decays in the heavy quark mass limit. This removes 
a major part of the theoretical uncertainty in the region of small 
invariant mass of the photon. We discuss in particular the 
decays  $B\to K^* \gamma$ and 
$B\to K^* \ell^+\ell^-$ and complete the calculation of 
corrections to the forward-backward asymmetry zero. The new 
correction shifts the asymmetry zero by 30\%, but the result 
confirms our previous conclusion that the asymmetry zero provides 
a clean phenomenological determination of the Wilson coefficient 
$C_9$.
\end{abstract}

\vskip 2.5cm

\centerline{\it (submitted to Nucl. Phys. B)}

\vfill

\end{titlepage}


\section{Introduction}

Radiative $B$ meson decays are interesting because they 
proceed entirely through loop effects. The chiral nature of weak 
interactions implies additional suppression factors thus 
enhancing the sensitivity of radiative decays to virtual effects 
in theories beyond the standard theory of flavour violation, if 
these new effects violate chirality. The observation of $b\to s \gamma$ 
transitions \cite{Ammar:1993sh} with a branching ratio of about $10^{-4}$ 
(in agreement with the standard theory) 
has since provided significant constraints on extensions of the 
standard model. The semi-leptonic decay $b\to s\ell^+\ell^-$ is 
expected to occur with a branching fraction of a few times $10^{-6}$ 
and will probably be first observed in the near future.

Radiative flavour-changing neutral current transitions can be 
measured inclusively over the hadronic final state (containing 
strangeness in the particular case considered here) or exclusively by 
tagging a particular light hadron, typically a kaon. The inclusive 
measurement is experimentally more difficult but theoretically 
simpler to interpret, since the decay rate is well and systematically 
approximated by the decay 
of a free $b$ quark into light quarks and gluons. In the long run, 
however, the easier detection of exclusive transitions requires the 
development of a systematic theoretical framework for these modes 
as well.  

The theoretical difficulty with exclusive decay modes 
is usually phrased as the need to know the hadronic form 
factors for the $B\to K^{(*)}$ transition, 
but this is only one aspect of the problem. 
Even if the form factors were known with infinite precision, the 
present treatment of exclusive, radiative decays would be incomplete, 
because there exist ``non-factorizable'' strong interaction effects 
that do not correspond to form factors. They arise from electromagnetic 
corrections to the matrix elements of purely hadronic operators in 
the weak effective Hamiltonian. We compute these non-factorizable 
corrections in this paper and demonstrate that exclusive, radiative 
decays can be treated in a similarly systematic manner as their inclusive 
counterparts. As a result we obtain the branching fractions for 
$B\to K^*\gamma$ and $B\to K^*\ell^+\ell^-$ for small invariant mass 
of the lepton pair to next-to-leading logarithmic (NLL) order in 
renormalization-group-improved perturbation theory. (In another 
commonly used terminology that takes into account the $1/\alpha_s$ 
enhancement of the semileptonic operator ${\cal O}_9$ 
the result would be 
referred to as next-to-next-to-leading logarithmic. However, in 
order not to have to distinguish in terminology the photonic 
and semileptonic decay, we shall refer to both as NLL when actually 
NNLL is meant for the semileptonic decay. Furthermore, since some 
of the 3-loop anomalous dimensions required for the semileptonic decay
are still not computed, the accuracy is not strictly NNLL.) 

Our method draws heavily on methods developed recently for other 
exclusive $B$ decays. In  \cite{Beneke:1999br,Beneke:2000ry} power 
counting in the heavy quark mass limit and standard factorization 
arguments for hard strong interaction processes have been used 
to demonstrate that decay amplitudes for hadronic two-body decays 
of $B$ mesons can be systematically computed in terms of 
form factors, light-cone distribution amplitudes and perturbative 
hard scattering kernels. A similar reasoning has been applied 
by two of us \cite{Beneke:2001wa} to hadronic form factors for 
$B$ decays into light mesons in the kinematic region of large recoil 
of the light meson. We also noted that a combination of the work 
on non-leptonic decays and on form factors could be used to compute 
non-factorizable corrections to radiative decay amplitudes. Furthermore, 
the ten different form factors that exclusive decays may depend on can 
be reduced to only three \cite{Charles:1999dr}. More 
precisely (but still schematically), the amplitude can be represented as 
\begin{equation}
\label{factform}
\langle \ell^+\ell^- \bar K^*_a|H_{\rm eff}|\bar{B}\rangle
= C_a \,\xi_a + \Phi_B\otimes T_a\otimes \Phi_{K^*},
\end{equation}
where $a=\perp,\parallel$ refers to a transversely and longitudinally 
polarized $K^*$, respectively. 
(An analogous result applies to the decay into a 
pseudoscalar meson, but we specify our notation to a vector meson for 
simplicity.) In this equation $\xi_a$ represent universal
heavy-to-light form factors \cite{Beneke:2001wa,Charles:1999dr} and 
$\Phi$ light-cone-distribution amplitudes. The factors 
$C_a$ and $T_a$ are calculable in renormalization-group improved 
perturbation theory. Previous treatments of exclusive radiative 
decays correspond to evaluating (\ref{factform}) to 
leading logarithmic accuracy (up to a weak annihilation
contribution as we discuss below). 
The main result of this paper is 
to show that systematic improvement is possible and to extend the 
accuracy to the next order. In Section~\ref{sec:nonfact} we present 
the result of the calculation. In Section~\ref{sec:analysis} we first 
discuss the inputs to our numerical result, such as Wilson coefficients and 
meson parameters. We then analyse the stability of the result with 
respect to the input parameters and a variation of the renormalization
scale. The phenomenological analysis is given in the subsequent two 
sections. Section~\ref{sec:gamma} is devoted to $B\to K^*\gamma$; 
Section~\ref{sec:ll} to $B\to K^*\ell^+\ell^-$. 
We focus in particular on corrections to the 
forward-backward asymmetry, which is independent on the form factors 
$\xi_a$ to first approximation. In the final Section~\ref{sec:final} 
we discuss power corrections related to photon structure on a 
qualitative level and summarize our conclusions.

\section{Non-factorizable corrections}
\label{sec:nonfact}

We generally neglect doubly Cabibbo-suppressed contributions to the 
decay amplitude. The weak effective Hamiltonian is then given by
\begin{equation}
\label{heff}
H_{\rm eff} = -\frac{G_F}{\sqrt{2}}\,V_{ts}^* V_{tb}\,
\sum_{i=1}^{10} C_i(\mu)\,{\cal O}_i(\mu),
\end{equation}
and we use the operator basis introduced in  \cite{Chetyrkin:1997vx} for 
the operators ${\cal O}_i$, $i=1,\ldots,6$. (More precisely, due to 
the normalization of (\ref{heff}), ${\cal O}_i=4 P_i$ with 
$P_i$ defined in  \cite{Chetyrkin:1997vx}.) We define 
\begin{equation}
{\cal O}_7=-\frac{g_{\rm em} \hat{m}_b}{8\pi^2}\,\bar{s}\sigma^{\mu\nu}
(1+\gamma_5)b F_{\mu\nu},\qquad
{\cal O}_8=-\frac{g_s \hat{m}_b}{8\pi^2}\,\bar{s}_i\sigma^{\mu\nu}
(1+\gamma_5)T^A_{ij}b_j G^A_{\mu\nu},
\end{equation}
\begin{equation}
{\cal O}_{9,10} = \frac{\alpha_{\rm em}}{2\pi}\,
(\bar{\ell} \ell)_{V,A}\,(\bar{s} b)_{\rm V-A},
\end{equation}
where $\alpha_{\rm em}=g_{\rm em}^2/{4\pi}$ is the fine-structure 
constant and $\hat{m}_b(\mu)$ the $b$ quark mass in the 
$\overline{\rm MS}$ scheme. We shall later trade the 
$\overline{\rm MS}$ mass for the pole mass $m_b$ and the 
PS mass $m_{b,\rm PS}$. To next-to-leading order
the pole and $\overline{\rm MS}$ masses are related by 
\begin{equation}
\label{polerel}
\hat{m}_b(\mu) = m_b\left(1+\frac{\alpha_s C_F}{4\pi}\left[
3\ln\frac{m_b^2}{\mu^2}-4\right]+O(\alpha_s^2)\right),
\end{equation}
with $\alpha_s\equiv\alpha_s(\mu)$. 
The sign convention for ${\cal O}_{7,8}$ corresponds 
to a negative $C_{7,8}$ and $+i g_s T^A$, $+i g_{\rm em} e_f$ 
for the ordinary quark--gauge-boson vertex ($e_f=-1$ for the lepton 
fields). We will present our result in terms 
of ``barred'' coefficients $\bar{C}_i$ (for  $i=1,\ldots,6$), 
defined as certain linear combinations 
of the $C_i$ as described in Appendix~\ref{app:a}. The linear
combinations are chosen such that the $\bar{C}_i$ coincide
{\em at leading logarithmic order} with 
the Wilson coefficients in the standard basis \cite{Buchalla:1996vs}.

\begin{figure}[t]
   \vspace{-6.9cm}
   \epsfysize=30cm
   \epsfxsize=22cm
   \centerline{\epsffile{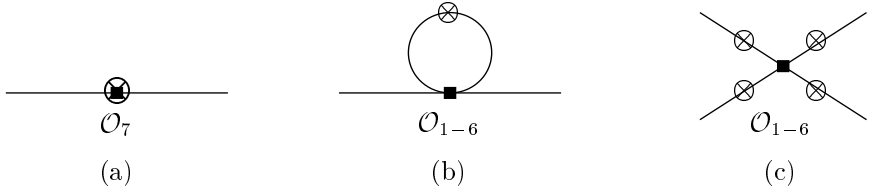}}
   \vspace*{-20.1cm}
\centerline{\parbox{14cm}{\caption{\label{fig2}
 Leading contributions to 
$\langle \gamma^*\bar K^*| H_{\rm eff} | \bar{B}\rangle$. 
The circled cross marks the possible insertions of the virtual 
photon line.}}}
\end{figure}

As for form factors and non-leptonic two-body decays there exist 
two distinct classes of non-factorizable effects. (By
``non-factorizable'' we mean all those corrections that are not 
contained in the definition of the QCD form factors for heavy-to-light
transitions. For example, the familiar leading-order diagrams 
shown in (a) and (b) of Figure~\ref{fig2} are factorizable.) 
The first class 
involves diagrams in which the spectator quark in the $B$ meson 
participates in the hard scattering. This effect occurs 
at leading order in an expansion in the strong coupling constant 
only through a weak annihilation diagram [Figure~\ref{fig2}c].
The relevant diagrams at next-to-leading order are shown 
as (a) and (b) in Figure~\ref{fig1} below and in Figure~\ref{fig3}. 
They contribute at order $\alpha_s^{0,1}$ to the functions $T_a$ in 
(\ref{factform}). Diagrams of this form have already been 
considered (for $q^2=0$) in \cite{Asatrian:1999mt}. However, 
bound state model wave-functions (rather than light-cone distribution 
amplitudes) were used and no attempt was made to systematically expand
the hard scattering amplitude in $1/m_b$. As a consequence, the result
of \cite{Asatrian:1999mt} for $\bar B\to K^*\gamma$ depends on an 
infrared cut-off. This difficulty is resolved in the present 
factorization approach. 
The second class contains all diagrams shown in the second 
row of Figure~\ref{fig1} below. Here the spectator quark is connected to the 
hard process represented by the diagram only through soft 
interactions. The result is therefore proportional to the form 
factor $\xi_a$ and the hard-scattering part 
gives an $\alpha_s$-correction to the functions 
$C_a$ in (\ref{factform}).

In this section we present the results of the calculation of these 
diagrams. 
Some of the results needed for diagrams of the second 
class can be extracted from work on inclusive radiative decays 
\cite{Greub:1996tg,Asatrian:2001de} 
and we have made use of these results as indicated below. 
The conventions for the form factors and light-cone 
distribution amplitudes for $B$ mesons and light mesons 
are those of  \cite{Beneke:2001wa}.

\subsection{Notation and leading-order result}

Since the matrix elements of the semi-leptonic operators 
${\cal O}_{9,10}$ can be expressed through 
$B\to K^*$ form factors, non-factorizable corrections contribute 
to the decay amplitude only through the production of a virtual 
photon, which then decays into the lepton pair. We therefore introduce
\begin{eqnarray}
\langle \gamma^*(q,\mu) \bar K^*(p',\varepsilon^*)| H_{\rm eff} | \bar{B}(p)
\rangle &=& -\frac{G_F}{\sqrt{2}}\,V_{ts}^* V_{tb}\,\frac{i g_{\rm em}
  m_b}{4\pi^2}
\Bigg\{2 \,{\cal T}_1(q^2) \,\epsilon^{\mu\nu\rho\sigma}
\varepsilon^{\ast}_\nu\, p_\rho p^{\prime}_\sigma\nonumber\\[0.0cm]
&& \hspace*{-4cm}
-i\,{\cal T}_2(q^2)\left[(M_B^2-m_{K^*}^2)\,\varepsilon^{\ast\mu}-
(\varepsilon^\ast\cdot
q)\,(p^\mu+p^{\prime\,\mu})\right]
\nonumber\\[0.0cm]
&& \hspace*{-4cm}
-i\,{\cal T}_3(q^2)\,(\varepsilon^\ast\cdot
q)\left[q^\mu-\frac{q^2}{M_B^2-m_{K^*}^2}(p^\mu+p^{\prime\,\mu})\right]
\Bigg\},
\label{caltdef}
\end{eqnarray}
where $H_{\rm eff}$ denotes the weak effective Hamiltonian. We define 
the overall quark mass factor as the pole mass here. The matrix
element decomposition is defined such that the leading order
contribution from the electromagnetic dipole operator ${\cal O}_7$ reads 
${\cal T}_i(q^2) = C_7 T_i(q^2)+\ldots$, where $T_i(q^2)$ denote the 
tensor form factors. Including also the four-quark operators (but
neglecting for the moment annihilation contributions), the 
leading logarithmic expressions are \cite{Grinstein:1989me}
\begin{eqnarray}
\label{firstT}
{\cal T}_1(q^2) &=& C_7^{\,\rm eff} \,T_1(q^2) + Y(q^2) \,\frac{q^2}
{2 m_b (M_B+m_{K^*})}\,V(q^2), \\
{\cal T}_2(q^2) &=& C_7^{\,\rm eff} \,T_2(q^2) + Y(q^2) \,\frac{q^2}
{2 m_b (M_B-m_{K^*})}\,A_1(q^2), \\
{\cal T}_3(q^2) &=& C_7^{\,\rm eff} \,T_3(q^2) + Y(q^2) \,\left[\frac{M_B-m_{K^*}}
{2 m_b} \,A_2
(q^2)- \frac{M_B+m_{K^*}}{2 m_b}\,A_1(q^2)\right], 
\label{lastT}
\end{eqnarray}
with $C_7^{\,\rm eff} = C_7-C_3/3-4 C_4/9-20 C_5/3-80 C_6/9=
C_7-(4 \bar{C}_3-\bar{C}_5)/9-(4 \bar{C}_4-\bar{C}_6)/3$ and 
\begin{eqnarray}
\label{yy}
Y(s) &=& h(s,m_c) \left(3 \bar{C}_1+\bar{C}_2+3 
\bar{C}_3+\bar{C}_4+3 \bar{C}_5+\bar{C}_6\right) 
\nonumber\\
&&-\,\frac{1}{2}\,h(s,m_b) \left(4 \,(\bar{C}_3+\bar{C}_4)+3 
\bar{C}_5+\bar{C}_6\right) 
-\frac{1}{2}\,h(s,0) \left(\bar{C}_3+3 \bar{C}_4\right)\nonumber\\
&&+\,\frac{2}{9}\,\left(\frac{2}{3}\bar{C}_3+2 \bar{C}_4+\frac{16}{3}  
\bar{C}_5\right).
\end{eqnarray}
The function
\begin{equation}
h(s,m_q) = -\frac{4}{9}\left(\ln\frac{m_q^2}{\mu^2} - \frac{2}{3} 
- z \right)- \frac{4}{9} \,(2+z) \,\sqrt{\,|z-1|} \,
\left\{
\begin{array}{l}
\,\arctan\displaystyle{\frac{1}{\sqrt{z-1}}}
\qquad\quad z>1\\[0.4cm]
\,\ln\displaystyle{\frac{1+\sqrt{1-z}}{\sqrt{z}}} - \frac{i\pi}{2}
\quad z\leq 1
\end{array}
\right.
\end{equation}
is related to the basic fermion loop. (Here $z$ is defined as $4 m_q^2/s$.) 
$Y(s)$ is given in the NDR scheme with anticommuting $\gamma_5$ and 
with respect to the operator basis of  \cite{Chetyrkin:1997vx}. 
Since $C_9$ is basis-dependent starting from next-to-leading
logarithmic order, the terms not proportional to $h(s,m_q)$ differ 
from those given in  \cite{Buchalla:1996vs}.
The contributions from the four-quark operators ${\cal O}_{1-6}$ are 
usually combined with the coefficient $C_9$ into an ``effective'' 
(basis- and scheme-independent) Wilson 
coefficient $C_9^{\,\rm eff}(q^2)=C_9+Y(q^2)$. 

The results of this 
paper are restricted to the kinematic region in which the energy 
of the final state meson scales with the heavy quark mass in the 
heavy quark limit. In practice we identify this with the region below the 
charm pair production threshold $q^2< 4 m_c^2 \approx 7\,\mbox{GeV}^2$. 
The various form factors appearing in (\ref{firstT})-(\ref{lastT}) 
are then related by symmetries \cite{Charles:1999dr,Beneke:2001wa}. 
Adopting the notation of  \cite{Beneke:2001wa}, 
(\ref{firstT})-(\ref{lastT}) simplify to
\begin{eqnarray}
\label{tperpdef}
&& {\cal T}_1(q^2) \equiv {\cal T}_\perp(q^2) = 
\xi_\perp(q^2) \left[C_7^{\,\rm eff} \,\delta_1 + 
\frac{q^2}{2 m_b M_B} \,Y(q^2)\right], \\
&&{\cal T}_2(q^2) = \frac{2 E}{M_B}\, {\cal T}_\perp(q^2), \\
&&{\cal T}_3(q^2) - \frac{M_B}{2 E}\,{\cal T}_2(q^2) \equiv  
{\cal T}_\parallel(q^2) = - \xi_\parallel(q^2) 
\left[C_7^{\,\rm eff} \,\delta_2 + \frac{M_B}{2 m_b}\, Y(q^2)
\,\delta_3\right],
\label{tpardef}
\end{eqnarray}
where $E=(M_B^2-q^2)/(2 M_B)$ refers to the energy of the 
final state meson and $\xi_{\perp,\parallel}$ refer to the form 
factors in the heavy quark and high energy limit. The factors 
$\delta_i$ are defined such that $\delta_i=1+O(\alpha_s)$.  
The $\alpha_s$-corrections have been computed in 
 \cite{Beneke:2001wa} and will be incorporated into the
next-to-leading order results later on.
The appearance of only two independent structures is a consequence of 
the chiral weak interactions and helicity conservation, and hence  
holds also after including next-to-leading order corrections 
\cite{Beneke:2001wa,Burdman:2001ku}. We therefore 
present our results in terms of the invariant amplitudes 
${\cal T}_{\perp,\,\parallel}(q^2)$, which refer to the decay into  
a transversely and longitudinally polarized vector meson 
(virtual photon), respectively. At next-to-leading order we represent 
these quantities in the form
\begin{eqnarray}
\label{nlodef}
{\cal T}_a &=& \xi_a \left(C_a^{(0)}+\frac{\alpha_s C_F}{4\pi} \,C_a^{(1)}
\right) 
\nonumber\\
&&\hspace*{0cm}+ \,\frac{\pi^2}{N_c}
\,\frac{f_B f_{K^*,\,a}}{M_B} \,\,\Xi_a\,
\sum_{\pm} \int\frac{d\omega}{\omega}\,\Phi_{B,\,\pm}(\omega)
\int_0^1 \!du\,\Phi_{K^*,\,a}(u) \,T_{a,\,\pm}(u,\omega),
\end{eqnarray}
where $C_F=4/3$, $N_c=3$, $\Xi_\perp\equiv 1$, $\Xi_\parallel
\equiv m_{K^*}/E$, and $T_{a,\,\pm}(u,\omega)$ is expanded as 
\begin{equation}
T_{a,\,\pm}(u,\omega)= T^{(0)}_{a,\,\pm}(u,\omega)+
\frac{\alpha_s C_F}{4\pi} \,T^{(1)}_{a,\,\pm}(u,\omega).
\end{equation}
$f_{K^*,\,\parallel}$ denotes the usual $K^*$ decay 
constant $f_{K^*}$. $f_{K^*,\,\perp}$ refers to the (scale-dependent) 
transverse decay constant defined by the matrix element of the tensor 
current. 
The leading-order coefficient $C^{(0)}_a$ follows by comparison with 
Eqs.~(\ref{tperpdef}) and (\ref{tpardef}) setting $\delta_i=1$.

To complete the leading-order result we have to compute the weak 
annihilation amplitude of Figure~\ref{fig2}c, which has no analogue 
in the inclusive decay and generates the hard-scattering term 
$T^{(0)}_{a,\,\pm}(u,\omega)$ in (\ref{nlodef}). To compute this 
term we perform the projection of the amplitude on the $B$ meson 
and $K^*$ meson distribution amplitude as explained in 
 \cite{Beneke:2001wa}. The four diagrams in Figure~\ref{fig2}c 
contribute at different powers in the $1/m_b$ expansion. It turns out 
that the leading contribution comes from the single diagram with the 
photon emitted from the spectator quark in the $B$ meson, because 
this allows the quark propagator to be off-shell by an amount 
of order $m_b\Lambda_{\rm QCD}$, the off-shellness being of order 
$m_b^2$ for the other three diagrams. With the convention that 
the $K^*$ meson momentum is nearly light-like in the minus light-cone 
direction, the amplitude for the surviving contribution depends only 
on the minus component of the spectator quark momentum. This is 
in contrast to the discussion in  \cite{Beneke:2001wa}, where, 
using the same convention for the outgoing meson, the hard-scattering 
amplitude depended only on the plus component of the spectator quark 
momentum. As a consequence the $B$ meson distribution amplitude 
$\Phi_{B,\pm}(\omega)$ now appears as coefficient of $\not\!\!n_{\mp}$ in the 
$B$ meson projector. Except for this difference the calculation 
follows the rules of  \cite{Beneke:2001wa}. The result reads
\begin{eqnarray}
\label{loann}
T^{(0)}_{\perp,\,+}(u,\omega) &=& T^{(0)}_{\perp,\,-}(u,\omega) = 
 T^{(0)}_{\parallel,\,+}(u,\omega) = 0\\
T^{(0)}_{\parallel,\,-}(u,\omega) &=& -e_q\,\frac{M_B\omega}{M_B \omega
-q^2-i\epsilon}\,\frac{4 M_B}{m_b} \,(\bar{C}_3+3 \bar{C}_4).
\end{eqnarray}
At leading order weak annihilation occurs only through penguin 
operators with small Wilson coefficients. For $B^\pm$ decay, 
there is an additional 
Cabibbo-suppressed contribution involving current-current 
operators, which can be obtained by multiplying the previous 
equation with the factor
\begin{equation}
-\frac{V_{us}^* V_{ub}}{V_{ts}^* V_{tb}}\cdot
\frac{\bar{C}_1+3 \bar{C}_2}{\bar{C}_3+3 \bar{C}_4},
\end{equation}
whose modulus is about 0.7. 
Both contributions are numerically small relative to the dominant 
leading-order terms. Note that since ${\cal T}_{\parallel}(q^2)$ 
does not contribute to the decay amplitude in the limit $q^2\to 0$, 
there is no weak annihilation contribution to $B\to K^*\gamma$ 
at leading order in $\alpha_s$ {\em and} at leading order in 
$1/m_b$.

We emphasize that our discussion of weak annihilation refers only 
to the leading contribution in the heavy quark limit. It is a novel 
aspect of $B\to K^*\ell^+\ell^-$ (with the $K^*$ longitudinally 
polarized) that weak annihilation does 
not vanish in this limit, if $q^2\sim m_b\Lambda_{\rm QCD}$. 
On the other hand, the sum of transverse and longitudinal
contributions is dominated by the transverse decay amplitude at small
$q^2$, and it is probable that the 
power-suppressed annihilation contributions to the transverse 
amplitude ${\cal T}_\perp(q^2)$, which we neglected, are numerically 
more important. For $q^2=0$ annihilation contributions of this 
sort have been studied in 
\cite{Grinstein:2000pc}. The interplay of the various terms 
merits consideration for $b\to d$ transitions, in which weak
annihilation through current-current operators is neither 
CKM- nor $\alpha_s$-suppressed. Since our main analysis concerns 
$b\to s$ decays, we defer this issue to a later discussion.

\subsection{Next-to-leading order -- spectator scattering}

\begin{figure}[t]
   \vspace{-3.5cm}
   \epsfysize=30cm
   \epsfxsize=22cm
   \centerline{\epsffile{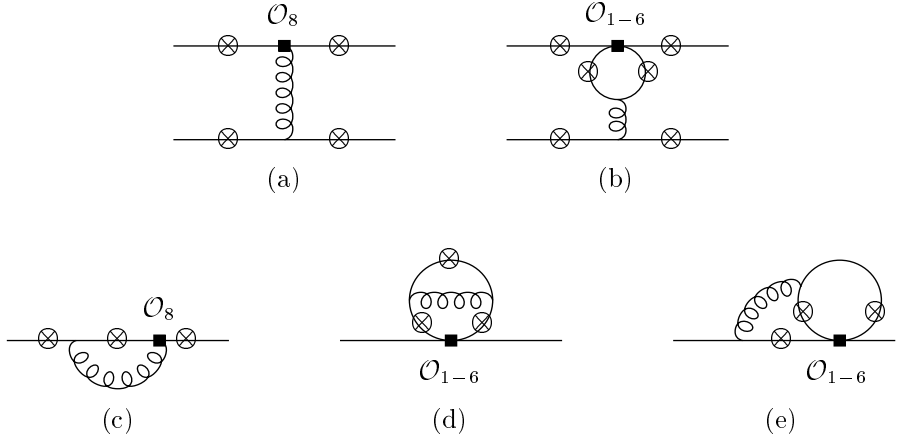}}
   \vspace*{-20.3cm}
\centerline{\parbox{14cm}{\caption{\label{fig1}
Non-factorizable contributions to 
$\langle \gamma^*\bar K^*| H_{\rm eff} | \bar{B}\rangle$. 
The circled cross marks the possible insertions of the virtual 
photon line. Diagrams 
that follow from (c) and (e) by symmetry are not shown. Upper line: 
hard spectator scattering. Lower line: diagrams involving a 
$B\to K^*$ form factor (the spectator quark line is not drawn for 
these diagrams).}}}
\end{figure}

The hard scattering functions $T^{(1)}_{a,\,\pm}$ in (\ref{nlodef}) 
contain a factorizable term from expressing the full QCD form factors 
in terms of $\xi_a$, related to the $\alpha_s$-correction to the 
$\delta_i$ in Eqs.~(\ref{tperpdef}), (\ref{tpardef}) above. 
We write $T^{(1)}_{a,\,\pm} = T_{a,\,\pm}^{(\rm f)}+ 
T_{a,\,\pm}^{(\rm nf)}$. The factorizable correction 
reads \cite{Beneke:2001wa}:
\begin{eqnarray}
\label{start1}
&& \hspace*{-0.5cm}
T^{(\rm f)}_{\perp,\,+}(u,\omega) = C_7^{\,\rm eff} \,
\frac{2 M_B}{\bar{u}E},\\
&& \hspace*{-0.5cm}
T^{(\rm f)}_{\parallel,\,+}(u,\omega) = \left[C_7^{\,\rm eff}  + 
\frac{q^2}{2 m_b M_B} \,Y(q^2)\right]\,
\frac{2 M_B^2}{\bar{u}E^2}
\\
&& \hspace*{-0.5cm}
T^{(\rm f)}_{\perp,\,-}(u,\omega) = 
T^{(\rm f)}_{\parallel,\,-}(u,\omega) = 0
\end{eqnarray}
The non-factorizable correction is obtained by computing matrix
elements of four-quark operators and the chromomagnetic dipole
operator represented by diagrams (a) and (b) in Figure~\ref{fig1}. The 
projection on the meson distribution amplitudes is straightforward. 
In the result we keep only the leading term in the heavy quark 
limit, expanding the amplitude in powers of the spectator quark 
momentum whenever this is permitted by power counting. In practice
this means keeping all terms that have one power of the spectator
quark momentum in the denominator. Such terms arise either from 
the gluon propagator that connects to the spectator quark line or 
from the spectator quark propagator, when the photon is emitted from 
the spectator quark line. We then find:
\begin{eqnarray}
T^{(\rm nf)}_{\perp,\,+}(u,\omega) &=&
-\frac{4 e_d \,C_8^{\,\rm eff}}{u+\bar{u} q^2/M_B^2} 
+\frac{M_B}{2 m_b}\,\Big[e_u t_\perp(u,m_c) \, (\bar{C}_2+\bar{C}_4-\bar{C}_6)
\nonumber\\
&& \hspace*{-1.5cm}+\,e_d \,t_\perp(u,m_b)\, 
(\bar{C}_3+\bar{C}_4-\bar{C}_6-4 m_b/M_B\,\bar{C}_5)+ e_d
\,t_\perp(u,0) 
\,\bar{C}_3\Big]
\label{Tnfperp}
\\
T^{(\rm nf)}_{\perp,\,-}(u,\omega) &=& 0
\\
T^{(\rm nf)}_{\parallel,\,+}(u,\omega) &=& \frac{M_B}{m_b}\,
\Big[e_u t_\parallel(u,m_c) \, 
(\bar{C}_2+\bar{C}_4-\bar{C}_6)+ e_d \,t_\parallel(u,m_b)\, 
(\bar{C}_3+\bar{C}_4-\bar{C}_6)
\nonumber\\
&& \hspace*{-1.5cm} + \,e_d \,t_\parallel(u,0) \,\bar{C}_3\Big]
\\
T^{(\rm nf)}_{\parallel,\,-}(u,\omega) &=& e_q\,\frac{M_B\omega}{M_B \omega
-q^2-i\epsilon}\, \bigg[\frac{8 \,C_8^{\,\rm eff}}{\bar{u}+u q^2/M_B^2} 
\nonumber\\
&& \hspace*{-1.5cm}
+\,\frac{6 M_B}{m_b}\, \Big(h(\bar{u}M_B^2+u q^2,m_c) 
\,(\bar{C}_2+\bar{C}_4+\bar{C}_6) 
+ h(\bar{u}M_B^2+u q^2,m_b)\,
(\bar{C}_3+\bar{C}_4+\bar{C}_6)
\nonumber\\
&& \hspace*{-1.5cm}
+ \,h(\bar{u}M_B^2+u q^2,0)\,
(\bar{C}_3+3 \bar{C}_4+3
\bar{C}_6)
-\frac{8}{27}\,(\bar{C}_3-\bar{C}_5-15\bar{C}_6)\Big)\bigg].
\label{tnf4}
\end{eqnarray}
Here $C_8^{\,\rm eff}=C_8+C_3-C_4/6+20 C_5-10 C_6/3 = 
C_8+(4 \bar{C}_3-\bar{C}_5)/3$, $e_u=2/3$, $e_d=-1/3$ and $e_q$ is 
the electric charge of the spectator quark in the $B$
meson. $h(s,m_q)$ has been defined above. The functions  
$t_a(u,m_q)$ arise from the two diagrams of Figure~\ref{fig1}b in which 
the photon attaches to the internal quark loop. They are given by 
\begin{eqnarray}
t_\perp(u,m_q) &=& \frac{2 M_B}{\bar{u} E} \,I_1(m_q) + 
\frac{q^2}{\bar{u}^2 E^2}\,(B_0(\bar{u} M_B^2+u q^2,m_q)-B_0(q^2,m_q))
\\
t_\parallel(u,m_q) &=& \frac{2 M_B}{\bar{u} E}\,I_1(m_q) + 
\frac{\bar{u} M_B^2+u q^2}{\bar{u}^2 E^2}\,
(B_0(\bar{u} M_B^2+u q^2,m_q)-B_0(q^2,m_q)),
\end{eqnarray}
where $B_0$ and $I_1$ are defined as
\begin{eqnarray}
\label{b0def}
&& \hspace*{-0.5cm}B_0(s,m_q)=-2\,\sqrt{4 m_q^2/s-1}\,\arctan\frac{1}
{\sqrt{4 m_q^2/s-1}}
\\
&& \hspace*{-0.5cm}I_1(m_q) = 1+\frac{2 m_q^2}{\bar{u} (M_B^2-q^2)}\,
\Big[L_1(x_+)+L_1(x_-)-L_1(y_+)-L_1(y_-)\Big],
\end{eqnarray}
and
\begin{eqnarray}
&& \hspace*{-0.5cm}x_\pm=\frac{1}{2}\pm\left(\frac{1}{4}-\frac{m_q^2}
{\bar{u} M_B^2+u q^2} \right)^{\!1/2},
\qquad
y_\pm=\frac{1}{2}\pm\left(\frac{1}{4}-\frac{m_q^2}{q^2}\right)^{\!1/2},
\\
&& \hspace*{-0.5cm}
L_1(x)=\ln\frac{x-1}{x}\,\ln(1-x)-\frac{\pi^2}{6}+\mbox{Li}_2\left(
\frac{x}{x-1}\right).
\end{eqnarray}
The correct imaginary parts are obtained by interpreting $m_q^2$ as 
$m_q^2-i\epsilon$. Closer inspection shows that contrary to 
appearance none of the hard-scattering functions is more singular than 
$1/\bar{u}$ as $u\to 1$. It follows that the convolution integrals 
with the kaon light-cone distribution are convergent at the endpoints. 

The limit $q^2\to 0$ ($E\to M_B/2$) 
of the transverse amplitude is relevant to the 
decay $\bar{B}\to \bar K^*\gamma$. The corresponding limiting function 
reads
\begin{equation}
t_\perp(u,m_q)_{|q^2=0} = \frac{4}{\bar{u}} \,\left(1+\frac{2 m_q^2}
{\bar{u} M_B^2}\,\Big[L_1(x_+)+L_1(x_-)\Big]_{|q^2=0}\right)
\end{equation}
In the 
same limit the longitudinal amplitude develops a logarithmic
singularity, which is of no consequence, because the longitudinal 
contribution to the $\bar{B}\to \bar K^*\ell^+\ell^-$ decay rate 
is suppressed by a power of $q^2$ relative to the transverse 
contribution in this limit. It does, however, imply that the 
longitudinal amplitude itself is sensitive to distances of order 
$1/\sqrt{q^2}$ and not perturbatively calculable unless $q^2\sim 
m_b\Lambda_{\rm QCD}$. (This long-distance sensitivity appears 
already at leading order in (\ref{tpardef}) through the light-quark 
contributions to $Y(q^2)$.)

\subsection{Next-to-leading order -- form factor correction}

The next-to-leading order coefficients 
$C^{(1)}_{a}$ in (\ref{nlodef}) 
contain a factorizable term from expressing the full QCD form factors 
in terms of $\xi_a$, related to the $\alpha_s$-correction to the 
$\delta_i$ in Eqs.~(\ref{tperpdef}), (\ref{tpardef}).  
We write $C^{(1)}_{a} = C_{a}^{(\rm f)}+ 
C_{a}^{(\rm nf)}$. The factorizable correction 
reads \cite{Beneke:2001wa}:
\begin{eqnarray}
\label{start2}
&& \hspace*{-0.5cm}
C^{(\rm f)}_{\perp} = C_7^{\,\rm eff} \left(4\ln\frac{m_b^2}{\mu^2} 
-4-L \right),\\
&& \hspace*{-0.5cm}
C^{(\rm f)}_{\parallel} = - C_7^{\,\rm eff} \left(4\ln\frac{m_b^2}{\mu^2} 
-6+4 L\right)
+ \frac{M_B}{2 m_b}\,Y(q^2)\,\Big(2-2 L\Big)
\label{end2}
\end{eqnarray}
with
\begin{equation}
\label{Ldef}
L\equiv -\frac{m_b^2-q^2}{q^2}\ln\left(1-\frac{q^2}{m_b^2}\right).
\end{equation}
The brackets multiplying $ C_7^{\,\rm eff}$ include the term 
$3\ln(m_b^2/\mu^2)-4$ from expressing the $\overline{\rm MS}$ quark 
mass in the definition of the operator ${\cal O}_7$ 
in terms of the $b$ quark pole
mass according to (\ref{polerel}). 
The non-factorizable correction is obtained by computing matrix
elements of four-quark operators and the chromomagnetic dipole
operator represented by diagrams (c) through (e) in 
Figure~\ref{fig1}. 

The matrix elements of four-quark operators require the calculation of
two-loop diagrams with several different mass scales. The result for
the current-current operators ${\cal O}_{1,2}$ is 
presented in  \cite{Asatrian:2001de} as a double expansion in 
$q^2/m_b^2$ and $m_c/m_b$. Since we are only interested in small 
$q^2$, this result is adequate for our purposes. (Note that only the 
result corresponding to Figure~1a-e of 
 \cite{Asatrian:2001de} is needed for our calculation.) The 2-loop 
matrix elements of penguin operators have not yet been computed and 
will hence be neglected. Due to the small Wilson coefficients of 
the penguin operators, this should be a very good approximation. 
The matrix element of the chromomagnetic dipole operator
[Figure~\ref{fig1}c] is also given in  \cite{Asatrian:2001de} in 
expanded form. The exact result is given in Appendix~\ref{app:b}. 
All this combined, we obtain
\begin{eqnarray}
\label{start3}
C_F C^{(\rm nf)}_{\perp} &=& -\,\bar{C}_2 F_2^{(7)}- C_8^{\rm eff}
F_8^{(7)}\nonumber\\
&&\hspace*{0.0cm}
-\,\frac{q^2}{2 m_b M_B} \left[\bar{C}_2 F_2^{(9)}+2\bar{C}_1\left(F_1^{(9)}+
\frac{1}{6}\,F_2^{(9)}\right)+C_8^{\rm eff} F_8^{(9)}\right] ,
\\
C_F C^{(\rm nf)}_{\parallel} &=& \bar{C}_2 F_2^{(7)}+C_8^{\rm eff}
F_8^{(7)}\nonumber\\
&&\hspace*{0.0cm}
+\,\frac{M_B}{2 m_b} 
\left[\bar{C}_2 F_2^{(9)}+2\bar{C}_1\left(F_1^{(9)}+
\frac{1}{6}\,F_2^{(9)}\right)+C_8^{\rm eff} F_8^{(9)}\right].
\label{end3}
\end{eqnarray}
The quantities $F_{1,2}^{(7,9)}$ can be extracted from 
 \cite{Asatrian:2001de}.\footnote{We thank M.~Walker for providing
  us with additional data points which cover the range $0.25\leq
  m_c/m_b\leq 0.35$ needed for the subsequent numerical analysis.} 
The quantities $F_{8}^{(7,9)}$ are 
given in Appendix~\ref{app:b} or can also be extracted from  
 \cite{Asatrian:2001de} in expanded form. 
In expressing the result in terms of the coefficients $\bar{C}_{1,2}$,
we have made use of $F_1^{(7)}+F_2^{(7)}/6=0$. We also substituted 
$C_8$ by $C_8^{\rm eff}$, taking into  account a subset of penguin
contributions.

\subsection{Weak annihilation}

\begin{figure}[t]
   \vspace{-3.8cm}
   \epsfysize=30cm
   \epsfxsize=22cm
   \centerline{\epsffile{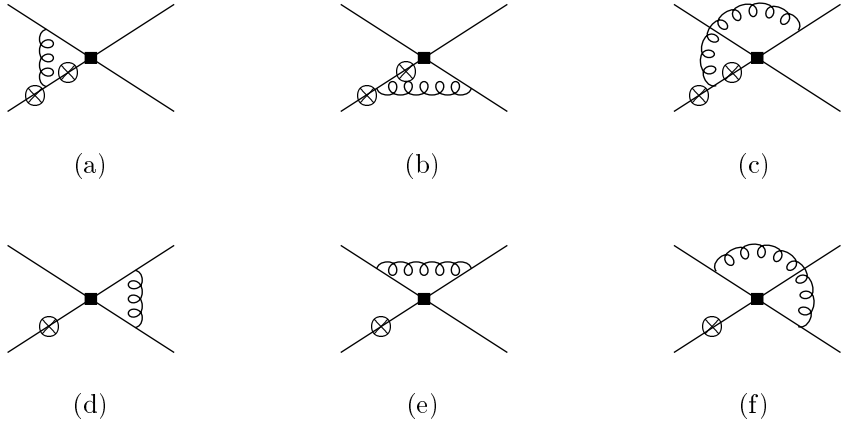}}
   \vspace*{-20.1cm}
\centerline{\parbox{14cm}{\caption{\label{fig3}
Vertex corrections to the 
weak annihilation diagram in Figure~\ref{fig2}c. Only the photon 
coupling to the spectator quark line contributes at leading 
order in $1/m_b$ as indicated by the circled cross.}}}
\end{figure}

In addition to the two sets of next-to-leading order corrections 
discussed up to now, there exist ``vertex corrections'' to the 
weak annihilation graph, as shown in Figure~\ref{fig3}. The distinction 
between weak annihilation and the previous spectator scattering
diagrams is not clear-cut. The diagrams in Figure~\ref{fig1}b with the 
photon attached to any of the external lines can also pass as a 
correction to the leading-order weak annihilation graph. In fact, 
the scale-dependent logarithm in $h(q^2,m_q)$ in (\ref{tnf4}) 
cancels part of the scale-dependence of the leading order 
annihilation amplitude, (\ref{loann}).

For $b\to s$ transitions current-current operators with large Wilson 
coefficients cannot be contracted to give diagrams of the type shown 
in Figure~\ref{fig3}. The total contribution is therefore suppressed by
three factors: the strong coupling constant; the small Wilson
coefficients of QCD penguin operators; and the suppression of the 
longitudinal amplitude in the decay rate for small $q^2$. We shall 
see subsequently that the leading-order weak annihilation effect 
is already small. Hence, although the radiative corrections to this 
leading-order term are of some conceptual interest in connection 
with renormalization of the $B$ meson distribution amplitude, we shall
not include the tiny radiative corrections represented in Figure~\ref{fig3} 
into our numerical analysis.

\subsection{Summary}

We now summarize our main result. The matrix element
$\langle \gamma^*(q,\mu) \bar K^*(p',\varepsilon^*)| H_{\rm eff} | \bar{B}(p)
\rangle$ is not scheme- and scale-independent in the standard operator 
formalism with an on-shell basis, unless the photon is
on-shell. Therefore ${\cal T}_{\perp,\,\parallel}(q^2)$ are not 
physical quantities. From the expressions for the decay rates 
given below, it will be clear that the following three 
quantities are independent of the conventions chosen to renormalize 
the weak effective Hamiltonian:
\begin{eqnarray}
\label{CC7}
{\cal C}_7 &\equiv& \frac{{\cal T}_{\perp}(0)}{\xi_\perp(0)} 
= C_7^{\rm eff} + \ldots\\
\label{CC9T}
{\cal C}_{9,\,\perp}(q^2) &\equiv& C_9+\frac{2 m_b M_B}{q^2}\,
 \frac{{\cal T}_{\perp}(q^2)}{\xi_\perp(q^2)}
= C_9+Y(q^2)+\frac{2 m_b M_B}{q^2}\, C_7^{\rm eff} + \ldots\\
{\cal C}_{9,\,\parallel}(q^2) &\equiv& C_9-\frac{2 m_b}{M_B}\,
 \frac{{\cal T}_{\parallel}(q^2)}{\xi_\parallel(q^2)}
= C_9+Y(q^2)+\frac{2 m_b}{M_B}\, C_7^{\rm eff} - e_q\,
\frac{4 M_B}{m_b}\,(\bar{C}_3+3\bar{C}_4)\,\nonumber\\
&& \hspace*{1cm}
\times\,\frac{\pi^2}{N_c}\,\frac{f_Bf_{K^*}}{M_B(E/m_{K^*})
\xi_\parallel(q^2)}
 \int d\omega\,\frac{M_B\Phi_{B,\,-}(\omega)}{M_B
   \omega-q^2-i\epsilon}+ 
  \ldots
\label{CC9L}
\end{eqnarray}
The quantity ${\cal C}_{9,\,\parallel}(q^2)$ depends on the charge of 
the $B$ meson through $e_q$, but this is suppressed in our notation. 
The ellipses denote the $\alpha_s$-corrections calculated above and
defined through (\ref{nlodef}). The explicit expressions for the 
quantities defined in (\ref{nlodef}) are given in 
(\ref{start2})-(\ref{end3}) and  (\ref{start1})-(\ref{tnf4}). We note 
that ${\cal C}_7$, ${\cal C}_{9,\perp}$ and ${\cal C}_{9,\parallel}$ 
depend on the conventions adopted in \cite{Beneke:2001wa} to define 
the soft form factors $\xi_\perp$ and $\xi_\parallel$. In addition 
${\cal C}_7$ depends also on the $b$ quark mass renormalization 
scheme (but $m_b \,{\cal C}_7$ does not). We discuss this further 
after (\ref{datastar}) below.

We verified that our results are scale-independent to the required 
order. More precisely, we have
\begin{equation}
\mu\frac{d}{d\mu}\left\{{\cal C}_7,{\cal C}_{9,\,\perp}(q^2),
{\cal C}_{9,\,\parallel}(q^2)\right\} = O(\alpha_s^2,\alpha_s C_{3-6}).
\end{equation}
The uncancelled terms at order $\alpha_s$ are proportional only to the
small Wilson coefficients of QCD penguin operators. They are related 
to the unknown two-loop matrix elements of penguin operators, and to 
the incomplete evaluation of weak annihilation effects in the case of 
${\cal C}_{9,\,\parallel}(q^2)$. Numerically, 
however, the missing contributions are negligible. When the strong  
interaction corrections to (\ref{CC7})-(\ref{CC9L}) are included, the 
next-to-leading logarithmic expression for $C_7^{\rm eff}$ 
\cite{Chetyrkin:1997vx} must be used. However, because $C_9\sim 
\ln(M_W/\mu)\sim 1/\alpha_s$ at leading order, the coefficient $C_9$ is needed 
to next-to-next-to-leading logarithmic order. The relevant 
initial conditions 
for the renormalization group evolution can be taken from 
\cite{Bobeth:2000mk} 
and we have incorporated them into our analysis. (The necessary 
renormalization group formulae are summarized in
Appendix~\ref{app:c}.) However, the three-loop anomalous dimension
matrix needed for the evolution is presently not known. Except for
this (and the numerically insignificant terms discussed above), our 
result for $\bar{B}\to \bar K^*\ell^+\ell^-$ is complete at 
next-to-next-to-leading logarithmic order.

\section{Numerical analysis of ${\cal C}_7$, 
${\cal C}_{9,\,\perp}$, ${\cal C}_{9,\,\parallel}$}
\label{sec:analysis}

\subsection{Specification of input parameters}

\label{input_sec}

{\em Wilson coefficients.} 
We begin our discussion of input parameters with the Wilson
coefficient $C_9$, which we need to next-to-next-to-leading logarithmic
order. In our analysis we use the complete 
next-to-next-to-leading logarithmic expression derived in
Appendix~\ref{app:c}, 
in which we set the unknown 3-loop anomalous dimensions 
$\kappa^{(1)}$ and $\gamma^{(2)}$ to zero. To estimate the 
uncertainty this introduces we note that the neglected terms read
\begin{equation}
\delta C_9 = -\,\frac{a_s(M_W)}{2\beta_0}\, 
\Bigg(\kappa^{(1)} \,D_{1}(\mu,M_W)+ \kappa^{(-1)}\,D_{-1}(\mu,M_W)\,
V\,\frac{[V^{-1} {\gamma^{(2)}}^T V]_{ij}}
{4\beta_0+\gamma_i^{(0)}-\gamma_j^{(0)}}\,V^{-1}\Bigg)_{\!2},
\end{equation}
where all notation is defined in Appendix~\ref{app:c} and the
subscript `2' refers to the component of the six-component vector 
in brackets. Under reasonable assumptions on the pattern of the 
unknown anomalous dimension matrix (vector) we observe that 
the largest contribution to the right hand side comes from 
$\kappa^{(1)}_2$. A rough estimate is therefore 
\begin{equation}
\delta C_9 \approx - 10^{-3} \,\kappa^{(1)}_2 \left[\left(
\frac{\alpha(\mu)}{\alpha(M_W)}\right)^{0.74}-1\right].
\end{equation}
Allowing $|\kappa^{(1)}_2|<100$, we 
conclude that at scales of order of the $b$ quark mass 
$C_9$ is known in the Standard Model to an accuracy of about 
$\pm 0.1$. (If $\kappa^{(1)}_2$ is much smaller than the upper limit
we assume, other terms may dominate $\delta C_9$, but the uncertainty 
estimate should remain valid.) This is to be compared to an 
error of $\pm 0.05$ due to the error on the 
top quark $\overline{\rm MS}$ mass,  
$\hat{m}_t(\hat{m}_t)=(167\pm5)\,$GeV.
The value of $C_9$ at the scale $\mu=4.6\,$GeV is given 
in Table~\ref{tab1} together with the remaining Wilson 
coefficients. The electroweak input parameters that go into these 
numbers are summarized along with other input parameters in 
Table~\ref{tab2} below. The strong coupling is always 
evolved according to the 3-loop formula.

\begin{table}[t]
\centerline{\parbox{14cm}{\caption{\label{tab1}
Wilson coefficients at the scale $\mu=4.6\,$GeV in leading-logarithmic
(LL) and next-to-leading-logarithmic order (NLL). Input parameters are 
$\Lambda^{(5)}_{\overline{\rm MS}}=0.220$\,GeV, $\hat m_t(\hat m_t)=167$\,GeV, 
$M_W=80.4$\,GeV, and $\sin^2\!\theta_W=0.23$. 3-loop running is used 
for $\alpha_s$. We also give $C_{9,10}$ at NNLL obtained as described
in the text.}}}
\vspace{0.1cm}
\begin{center}
\begin{tabular}{|l|c|c|c|c|c|c|}
\hline\hline
\rule[-2mm]{0mm}{7mm}
 & $\bar{C}_1$ & $\bar{C}_2$ & $\bar{C}_3$ & $\bar{C}_4$ & $\bar{C}_5$
 & $\bar{C}_6$ \\
\hline
\rule[-0mm]{0mm}{4mm}
LL    & $-0.257$ & $1.112$ & $0.012$ & $-0.026$ & $0.008$ & $-0.033$ \\

NLL   & $-0.151$ & $1.059$ & $0.012$ & $-0.034$ & $0.010$ & $-0.040$ \\
\hline
\rule[-2mm]{0mm}{7mm}
 & $C_7^{\rm eff}$ & $C_8^{\rm eff}$ & $C_9$ & $C_{10}$
 & $C_9^{\rm NNLL}$ &  $C_{10}^{\rm NNLL}$ \\
\hline
\rule[-0mm]{0mm}{4mm}
LL  & $-0.314$ & $-0.149$ & $2.007$ & 0
 & & \\
NLL & $-0.308$ & $-0.169$ & $4.154$ & $-4.261$
 & \raisebox{2.5mm}[-2.5mm]{$4.214$} & \raisebox{2.5mm}[-2.5mm]{$-4.312$} \\
\hline\hline
\end{tabular}
\end{center}
\end{table}

\vskip0.2cm\noindent
{\em Quark masses.} Up to now we have assumed that $m_b$ is the 
pole mass. It is usually assumed that the pole mass of the $b$ quark 
is less well known than the $\overline{\rm MS}$ mass or the 
potential-subtracted (PS) mass \cite{Beneke:1998rk}. (The PS mass 
replaces the pole mass in quantities in which the $b$ quark is nearly 
on-shell, but has some large infrared contributions removed.) We 
therefore replace $m_b$ by the PS mass through the relation 
\begin{equation}
\label{polePS}
m_b= m_{b,\rm PS}(\mu_f) + \frac{4\alpha_s}{3\pi}\,\mu_f
\end{equation}
and use $m_{b,\rm PS}(2\,\mbox{GeV})=(4.6\pm 0.1)\,$GeV 
\cite{Beneke:1999fe} as an input parameter. An exception to this 
replacement is applied to the function $Y(s)$ in which we keep the 
$b$ quark pole mass -- computed according to (\ref{polePS}) -- 
in the small contributions from $b$ quark loops. Apart from
reinterpreting $m_b$ as $m_{b,\rm PS}(2\,\mbox{GeV})$ the only 
consequence of introducing the PS mass is an additional term 
$4\mu_f/m_b$ that appears in the round brackets multiplying 
$C_7^{\rm eff}$ in (\ref{start2}), (\ref{end2}). We use the 
charm quark pole mass, $m_c=(1.4\pm 0.2)\,$GeV.

\vskip0.2cm\noindent
{\em Meson parameters.} Many of the relevant meson parameters are 
not directly known from experiment. QCD sum rule calculations 
substitute for this lack of information and we use 
\cite{Ball:1998kk} as our source of information. 
A summary of all input parameters together with their assumed errors 
is given in Table~\ref{tab2}. We now
discuss some of the input parameters in more detail. 

 \begin{table}[t] 
\centerline{\parbox{14cm}{\caption{\label{tab2}
Summary of input parameters and estimated uncertainties.}}}
\vspace{0.1cm}
   \begin{center} 
     \begin{tabular}{| l l| l l |} 
\hline  
\hline 
\rule[-2mm]{0mm}{7mm}
     $M_W$                           & $80.4$~GeV & 
        $f_B$                           & $180 \pm 30$~MeV\\ 
     $\hat m_t(\hat m_t)$            & $167 \pm 5$~GeV & 
        $\lambda_{B,+}^{-1}$            & $(3 \pm 1)$~GeV$^{-1}$ \\
     $|V_{ts}|$                      & $0.041 \pm 0.003$ &
        $f_{K^*,\perp}(1\,\mbox{GeV})$  & $185 \pm 10$~MeV \\
     $\alpha_{\rm em}$               & $ 1/137$ &
        $f_{K^*,\parallel}$             & $225 \pm 30$~MeV \\ 
     $\Lambda_{\rm QCD}^{(n_f=5)}$   & $220 \pm 40$~MeV &
       $a_1(\bar K^*)_{\perp,\,\parallel}$    & $0.2 \pm 0.1$ \\ 
     $m_{b,\rm PS}(2\,\mbox{GeV})$   & $4.6 \pm 0.1$~GeV &
       $a_2(\bar K^*)_{\perp,\,\parallel}$    & $0.05 \pm 0.1$  \\
     $m_c$                           & $1.4 \pm 0.2$~GeV &
     $M_B\,\xi_\parallel(0)/(2 m_{K^*})$   & $ 0.47 \pm 0.09$ \\
     & & $\xi_\perp(0)$                & $ 0.35 \pm 0.07$ 
\\[0.15cm]
\hline\hline
       \end{tabular} 
\end{center} 
\end{table} 

By convention \cite{Beneke:2001wa} 
the soft form factors (including the one for a
pseudoscalar kaon) at zero momentum transfer 
are related to the full form factors by 
\begin{eqnarray}
\label{ins}
\xi_P(0) = F_+^K(0),\quad
\xi_\perp(0) = \frac{M_B}{M_B+m_{K^*}}\,V^{K^*}(0),\quad
\frac{M_B}{2 m_{K^*}}\,\xi_\parallel(0) = A_0^{K^*}(0) .
\end{eqnarray}
to all orders in perturbation theory. 
For $A_0^{K^*}(0)$ we have taken the QCD sum rule
result from \cite{Ball:1998kk} and this gives the soft form factor 
$\xi_\parallel(0)$  
listed in Table~\ref{tab2}. The choice $\xi_\perp(0) = 0.35 \pm 0.07$,
also given in that Table, requires comment since (\ref{ins}) and 
$V^{K^*}(0)$ from \cite{Ball:1998kk} would give 
$\xi_\perp(0) = 0.39\pm 0.06$. The motivation for our choice of 
input is related to the fact that the QCD sum rule 
prediction for $T_1^{K^*}(0)/V^{K^*}(0)$ deviates from the behaviour 
expected in the heavy quark limit \cite{Beneke:2001wa}. An alternative
way of computing $\xi_\perp(0)$ uses $\xi_\perp(0)\approx 0.93\, 
T_1^{K^*}(0)$ (with $T_1^{K^*}(0)$ evaluated at the scale $\mu=m_b$). 
Taking $T_1^{K^*}(0)$ from \cite{Ball:1998kk} gives
$\xi_\perp(0)\approx 0.35$. Using this smaller input has the 
advantage that our result for the $\bar B\to\bar K^*\gamma$ decay 
rate is automatically consistent with an alternative representation of
this decay rate in terms of the full QCD form factor $T_1^{K^*}(0)$. 
(This will be discussed in more detail later.) Furthermore, we shall 
see that the decay rate predicted for $\bar B\to\bar K^*\gamma$ 
at next-to-leading order favours form factors smaller than those 
estimated with QCD sum rules. We may then take the point of view that 
$\xi_\perp(0)$ is {\em determined} by 
$\Gamma(\bar B\to\bar K^*\gamma)$ and hence also  $T_1^{K^*}(0)$ 
and $V^{K^*}(0)$ through the relations of \cite{Beneke:2001wa}. 
This logic of course implies that we take seriously the heavy 
quark limit prediction for the decay rate and form factor 
relations in the heavy quark limit. 
The energy dependence of the form factors is 
assumed to be 
\begin{equation}
\xi_\perp(q^2) = \xi_\perp(0)
\left(\frac{1}{1-q^2/M_B^2}\right)^{\!2},
\qquad
\xi_\parallel(q^2) = \xi_\parallel(0) \left(\frac{1}{1-q^2/M_B^2}\right)^{\!3},
\end{equation}
as predicted by power counting in the heavy quark limit. 

The hard-spectator scattering (and annihilation) contribution 
  depend on the light-cone wave functions of the $B$-meson and the 
  light meson in the final state which are characterized in terms of 
  decay constants, Gegenbauer coefficients etc.  
  The leading contribution in the heavy quark mass expansion involves 
  the leading-twist distribution amplitudes of light mesons only. 
  We truncate the expansion of these functions into  
  Gegenbauer-polynomials $C_i^{(3/2)}$ at second order, i.e.\  
  ($a=\perp,\parallel$) 
  \begin{equation} 
    \Phi_{\bar K^*,a}(u) = 6 u (1-u)  
      \left\{ 1 + a_1(\bar K^*)_a \, C_1^{(3/2)}(2u-1)  
               + a_2(\bar K^*)_a \, C_2^{(3/2)}(2u-1) \right\} \ . 
  \end{equation} 
  The values for the Gegenbauer coefficients   
  are taken from  \cite{Ball:1998kk}, but for
  simplicity we neglect the small differences between 
  $a_i(\bar K^*)_\perp$ and 
  $a_i(\bar K^*)_\parallel$. We also enlarge the error, see Table~\ref{tab2}. 
  In the same Table we also give numerical values for the light 
  meson decay constants \cite{Ball:1998kk}.  
 (The Gegenbauer coefficients and the decay constant for a
  transversely 
  polarized vector meson, $f_\perp$, are scale-dependent and assumed to be 
  evaluated at the scale~1~GeV. The variation of Gegenbauer moments
  has a negligible effect on our result compared to the overall 
parameter uncertainties. We therefore neglect the scale-dependence 
of the Gegenbauer moments in the numerical analysis, but we evolve 
  $f_\perp$ according to $f_\perp(\mu) = f_\perp(\mu_0)  
  \,\left(\alpha_s(\mu)/\alpha_s(\mu_0)\right)^{4/23}$.) The
  renormalization scale in the hard-scattering and annihilation terms 
is chosen lower than in the form factor contributions in order to 
reflect the fact that the typical virtualities in the hard scattering
term are of order $m_b\Lambda_{\rm QCD}$ rather than $m_b^2$. If
$\mu_1$ (assumed to be of order $m_b$) is the scale in the form factor
term, we choose $(\mu_1\Lambda_h)^{1/2}$ with $\Lambda_h=0.5\,$GeV in 
 the hard-scattering term. This convention applies to all
 scale-dependent quantities in the hard-scattering term including 
$\alpha_s$ and the Wilson coefficients.

The two $B$ meson light-cone distribution amplitudes enter only through 
the two ``moments''
  \begin{eqnarray} 
    \lambda_{B,+}^{-1} 
    &=& \int_0^\infty d\omega \,  
      \frac{\Phi_{B,+}(\omega)}{\omega} \ ,  
    \label{Bmomdef1}\\ 
    \lambda_{B,-}^{-1}(q^2) 
    &=& \int_0^\infty d\omega \,  
      \frac{\Phi_{B,-}(\omega)}{\omega-q^2/M_B-i\epsilon} 
  \label{Bmomdef2} 
  \end{eqnarray}  
The moment of $\Phi_{B,+}$ is identical to the moment that appears
also in non-leptonic $B$ decays \cite{Beneke:1999br}, the decay 
$B\to \gamma l\nu$ \cite{Korchemsky:2000qb} and in the factorization 
of form factors \cite{Beneke:2001wa}. The appearance of  $\Phi_{B,-}$ 
at leading order in the heavy quark expansion is a new aspect of 
the decay $\bar{B}\to \bar K^*\ell^+\ell^-$.

\begin{figure}[t] 
        \begin{center} 
         \unitlength1truecm 
         \begin{picture}(7.5,4.5) 
          \put(0.2,0){\psfig{file=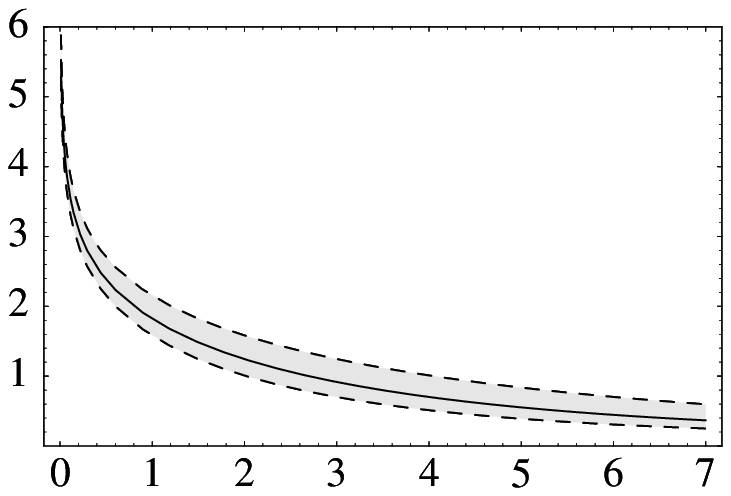,width = 7truecm}} 
          \put(3,-0.3){{\small $q^2\ [{\rm GeV}^2]$}} 
          \put(-1.6,2.2){\small $\displaystyle 
          \frac{|\lambda_{B,-}^{-1}(q^2)|}{\lambda_{B,+}^{-1}}$} 
          \end{picture} 
        \end{center} 
\centerline{\parbox{14cm}{\caption{  \label{Bminusmomfig} 
The absolute value of the ratio of 
            $B$-meson moments $|\lambda_{B,-}^{-1}(q^2)|/\lambda_{B,+}^{-1}$ 
            as a function of $q^2$. 
            The $B$-meson distribution amplitudes are taken 
            as in (\ref{GN_WF}) with 
$\omega_0^{-1}=(3 \pm 1)$~GeV${}^{-1}$.}}}
\end{figure} 
  
The light-cone distribution amplitudes $\Phi_{B,\pm}(\omega)$ 
appearing in (\ref{Bmomdef1}), (\ref{Bmomdef2}) are not
well-constrained at present and provide a 
  major source of uncertainty in our calculation.
Some general properties have been derived in 
  the literature \cite{Beneke:2001wa,Grozin:1997pq}.  
  The equations of motions for the light quark in  the $B$-meson 
  lead to 
  \begin{equation} 
       \Phi_{B,-}(\omega) =  
       \int_0^1 \frac{d\eta}{\eta} \, \Phi_{B,+}(\omega/\eta) 
   \quad \Leftrightarrow \quad 
       \Phi_{B,+}(\omega) = - \omega \, \Phi'_{B,-}(\omega) 
   \label{Brel} 
   \end{equation} 
We conclude from this that $\lambda_{B,+}^{-1} = \Phi_{B,-}(0)$. 
Furthermore $\lambda_{B,-}^{-1}(q^2)$ must diverge logarithmically for 
   $q^2 \to 0$, provided that $\lambda_{B,+}^{-1} \neq 0$: 
    \begin{eqnarray} 
    \lim_{q^2 \to 0} \lambda_{B,-}^{-1}(q^2) = 
      \lambda_{B,+}^{-1} \, \int_0^1 \frac{d\eta}{\eta} \to \infty 
    \label{Bmomentlog} 
    \end{eqnarray} 
     On the other hand, for $q^2 = O(M_B \Lambda_{\rm QCD})$  
     the moment $\lambda_{B,-}^{-1}(q^2)$ is finite and of 
     order~$1/\Lambda_{\rm QCD}$. 
    The equations of motion for the 
    heavy quark relate the first moments $\langle \omega \rangle_{B,\pm}$  
    to the mass difference $\bar \Lambda_{\rm HQET} = M_B - m_b$, 
    leading to $\langle \omega \rangle_{B,+} =  
    2 \langle \omega \rangle_{B,-} = 4\bar 
    \Lambda_{\rm HQET}/3$. 
     For  $\lambda_{B,+}$ the upper bound  
     $4\bar \Lambda_{\rm HQET}/3$ has been derived in  
     \cite{Korchemsky:2000qb}. 
  The simple model functions  
  \begin{equation} 
     \Phi_{B,+}(\omega) = \frac{\omega}{\omega_0^2} \, 
     e^{-\omega/\omega_0} \ , \qquad 
      \Phi_{B,-}(\omega) = \frac{1}{\omega_0} \, 
     e^{-\omega/\omega_0}  
  \label{GN_WF} 
  \end{equation} 
  with $\omega_0 = 2\bar 
  \Lambda_{\rm HQET}/3$, that have been proposed 
  in  \cite{Grozin:1997pq}, are consistent with  
  these general constraints, leading to 
  $\lambda_{B,+} = 2 \bar \Lambda_{\rm HQET}/3$, and 
  \begin{equation} 
  \lambda_{B,-}^{-1}(q^2) =  
   \frac{e^{-q^2/(M_B\omega_0)}}{\omega_0}  \, 
     \left[ - {\rm Ei}(q^2/M_B\omega_0) + i \pi \right] 
  \label{Bmomentexample} 
  \end{equation} 
  where ${\rm Ei}(z)$ is the exponential integral function. 
   Numerically, for $\bar \Lambda_{\rm HQET} \simeq 500$~MeV, 
   one obtains $\lambda_{B,+}^{-1} \simeq 3$~GeV$^{-1}$ with an estimated  
   error of about $1$~GeV$^{-1}$. For the same parameter values  
   the absolute value of  
   $\lambda_{B,-}^{-1}(q^2)$, 
   normalized to $\lambda_{B,+}^{-1}$ is plotted in 
   Figure~\ref{Bminusmomfig}. 
   There exist alternative proposals for $B$-meson light-cone 
   wave functions, for instance from the Bauer-Stech-Wirbel 
   model \cite{Wirbel:1985ji} or variants of it. Since 
   in these models the distinction between $\Phi_{B,+}(\omega)$ and 
   $\Phi_{B,-}(\omega)$ is not made,   
   we refrain from presenting a thorough comparison of different 
   models and stick to (\ref{Bmomentexample}) to evaluate the moment 
of $\Phi_{B,-}(\omega)$. This means that we neglect a 
   systematic uncertainty related to the {\em shape}\/ of 
   $\Phi_{B,-}(\omega)$, but this uncertainty is irrelevant
   numerically, because the moment in question appears only in the
   small annihilation contribution and the small correction to 
   ${\cal C}_{9,\,\parallel}(q^2)$.

   All input values from the meson sector together with their 
   estimated uncertainties are summarized in Table~\ref{tab2}. 
   We note that apart from the renormalization scale uncertainty 
and the error in our knowledge of $\alpha_s$, the most important 
uncertain parameters can be collected into a single factor 
\begin{equation}
\label{combpar}
    \frac{\pi^2 \, f_B  \, f_{K^*,\,a}} 
          {N_c M_B \, \lambda_{B,+}\,\xi_a(0)}
\end{equation}
that determines the relative magnitude of the hard-scattering versus 
the form factor term. Adding all errors in quadrature, this factor 
is uncertain by about $\pm 50\%$, where the largest error is 
currently from $\lambda_{B,+}^{-1}$.  

\subsection{Exclusive effective ``Wilson'' coefficients}
\label{sec:output}

Having specified our numerical input, we now discuss the three 
effective ``Wilson'' coefficients ${\cal C}_7$, 
${\cal C}_{9,\,\perp}(q^2)$ and ${\cal C}_{9,\,\parallel}(q^2)$. 

\begin{figure}[t]
\vspace*{3.6cm}
\hspace*{-0cm}
\epsffile{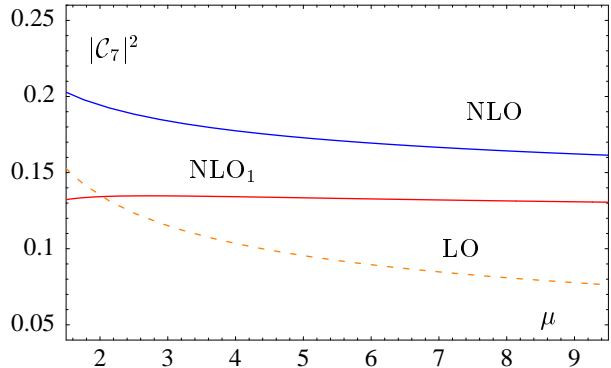}
\vskip0.1cm
\centerline{\parbox{14cm}{\caption{\label{fig:c7fig}
Renormalization scale-dependence of $|{\cal C}_7|^2$ at leading 
(LO) and next-to-leading order (NLO). The curve ``NLO${}_1$'' shows 
the NLO result without the spectator scattering correction.}}}
\end{figure}

We begin with the quantity $|{\cal C}_7|^2$ to which the decay rate 
of $\bar{B}\to \bar K^*\gamma$ is proportional. In Figure~\ref{fig:c7fig} 
we show the renormalization scale dependence of this quantity at 
leading and at next-to-leading order. We also show a curve that 
corresponds to setting the hard-scattering term to zero, i.e. to 
taking into account only the correction $C_a^{(1)}$ in
(\ref{nlodef}). The reason for considering this term separately is 
that it should cancel the sizeable leading-order scale dependence, 
while the hard scattering correction is a physically different 
effect that appears first at next-to-leading order. 
Figure~\ref{fig:c7fig} shows that this is indeed correct. The 
hard scattering correction reintroduces a mild scale-dependence. The 
most important effect is however a large enhancement of 
$|{\cal C}_7|^2$ at next-to-leading order. At the scale 
$m_b=4.6\,$GeV we find
\begin{equation}
|{\cal C}_7|^2_{\rm NLO}/|{\cal C}_7|^2_{\rm LO} \approx 1.78,
\end{equation}
which corresponds to a sizeable, but not unreasonable $33\%$ 
correction on the amplitude level. 
The form factor and hard-scattering correction contribute about 
equally to this enhancement. More precisely, the [non-]factorizable 
part of $C_\perp^{(1)}$ (defined in (\ref{nlodef})) 
is a $-8\%$ [$+24\%$] correction to the real part of the 
amplitude, the  [non-]factorizable part of $T_{\perp,\,+}^{(1)}$ 
is a $+11\%$ [$+5\%$] correction (all numbers at $\mu=m_b$).
The error on $|{\cal C}_7|^2$ is 
estimated by combining in quadrature the error from all input 
parameters as specified in the previous section and from allowing the 
renormalization scale to vary from $m_b/2$ to $2 m_b$. (The scale in 
the hard scattering term is accordingly lower, see the discussion 
above.). The result is
\begin{equation}
|{\cal C}_7|^2_{\rm NLO} = 0.175^{+0.029}_{-0.026},
\end{equation}
where the largest errors are from scale dependence ($\pm 0.014$), 
$\lambda_{B,+}^{-1}$ ($\pm 0.015$), $\xi_\perp(0)$ ($\pm 0.009$), 
and $\Lambda_{\rm QCD}$ 
($\pm 0.010$). The value given refers to using the PS mass 
$m_{b,\rm PS}(2\,\mbox{GeV})$, since only the product 
$m_{b,\rm PS}(2\,\mbox{GeV})\,|{\cal C}_7|$ is
convention-independent. The NLO result for $|{\cal C}_7|^2$ must 
be compared to $|{\cal C}_7|^2_{\rm LO} = 0.098$ with a $30\%$ error 
from scale dependence alone.
At this place it is also appropriate to note that 
our result is based on the heavy quark expansion, and therefore 
there exist corrections of order $\Lambda_{\rm QCD}/m_b$ at the 
amplitude level. At present we have no means of quantifying these 
corrections in a systematic way.

\begin{figure}[p]
\vspace*{2cm}
\hspace*{-4cm}
\epsffile{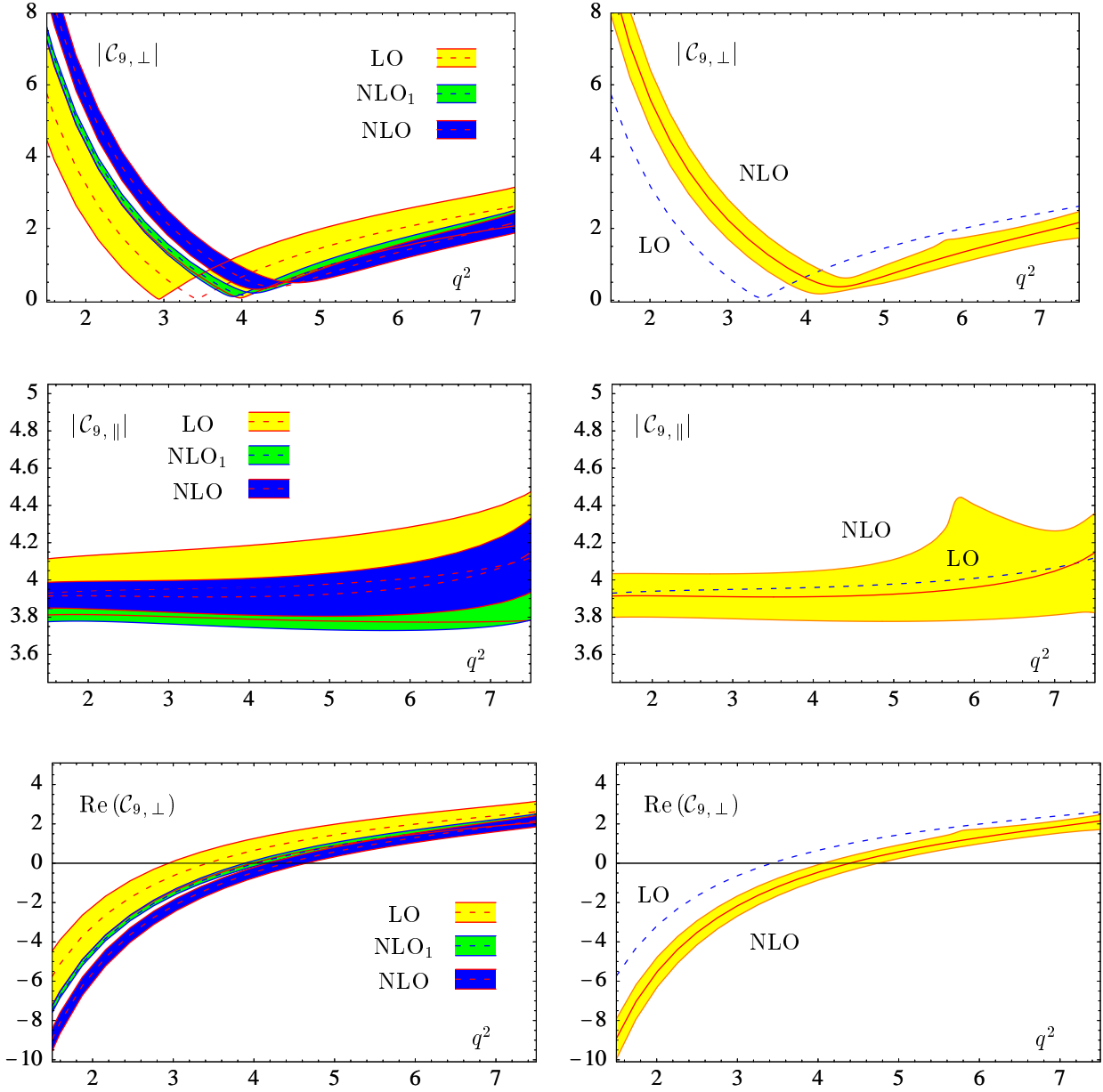}
\vskip0.5cm
\centerline{\parbox{14cm}{\caption{\label{fig:c9fig}
Momentum-transfer dependence of $|{\cal C}_{9,\,\perp}|$, 
$|{\cal C}_{9,\,\parallel}|$ and $\mbox{Re}\,({\cal C}_{9,\,\perp})$. 
The left panels show the LO, NLO and NLO result without spectator 
scattering (``NLO${}_1$'') for fixed input parameters. The width of 
the bands demonstrates the renormalization scale dependence estimated 
by variation from $m_b/2$ to $2 m_b$. The right panel shows the 
next-to-leading order result including renormalization scale and 
input parameter uncertainties (all added in quadrature), and for 
comparison the leading order result (dashed). The solid line in the
center of the band refers to the NLO result with default parameters. 
$|{\cal C}_{9,\,\parallel}|$ is given for $B^-$ mesons.}}}
\end{figure}

The $q^2$-dependent coefficients ${\cal C}_{9,\,\perp}(q^2)$, 
${\cal C}_{9,\,\parallel}(q^2)$ defined in (\ref{CC9T}), (\ref{CC9L}) 
are shown in Figure~\ref{fig:c9fig}. The left panels display the 
reduction of renormalization scale dependence in going from leading 
order to next-to-leading order (including the intermediate
approximation with the hard scattering term set to zero). The 
right panel gives the complete NLO result with all input parameter 
uncertainties and scale dependence included in the error estimate. 

The characteristic features of the result can be understood from 
the discussion of ${\cal C}_7$ above. Since $C_9>0$ and $C_7^{\rm
  eff}<0$, the contribution to  ${\cal C}_{9,a}(q^2)$ 
from the virtual photon matrix element, ${\cal T}_a(q^2)$, is 
always negative ($a=\perp,\parallel$). In the case of a transverse 
virtual photon, this contribution is enhanced by a factor 
$M_B^2/q^2$ owing to the real photon pole and hence dominates 
${\cal C}_{9,\,\perp}$ at small $q^2$. This causes 
$\mbox{Re}\,({\cal C}_{9,\,\perp})$ to change its sign and 
results in the characteristic shape of $|{\cal C}_{9,\,\perp}|$. 
Due to the large contribution from the photon pole the next-to-leading
order contribution is again substantial and the value of momentum 
transfer $q_0^2$ at which $\mbox{Re}\,({\cal C}_{9,\,\perp})$ changes its 
sign is increased. Since $q_0^2$ determines the location 
of the forward-backward asymmetry zero, this important correction 
will be discussed in some detail in Section~\ref{sec:AFB}.  
The photon pole is absent in the longitudinal 
coefficient ${\cal C}_{9,\,\parallel}$, so the NLO correction to it 
is much smaller. The form
factor and hard scattering correction (both small) nearly compensate
each other leaving no net modification of the leading order 
coefficient. Contrary to the transverse coefficient, the longitudinal
one depends on the charge of the spectator quark in the $B$ meson,
i.e. on the charge of the decaying $B$ meson. This effect is already
present at leading order via the annihilation contribution. At NLO 
it amounts to a few percent of ${\cal C}_{9,\,\parallel}$, the precise 
magnitude being $q^2$-dependent. We return to a discussion of this 
isospin-breaking effect in the context of the $\bar{B}\to\bar{K}^* 
\ell^+\ell^-$ decay spectrum.

The peculiar discontinuity in the error band 
around $q^2=6\,\mbox{GeV}^2$ in two 
of the right panels of Figure~\ref{fig:c9fig} is unphysical and 
related to the fact that in computing the error we allow the charm 
quark mass to be as small as $1.2\,$GeV. The discontinuity is a 
consequence of the $c\bar{c}$ threshold in the charm loop diagrams 
and reminds us that the validity of our calculation is not only 
limited by the requirement that the $K^*$ momentum is large (which
puts an upper bound on $q^2$), but also by the lack of a
model-independent treatment of the resonant charmonium 
contributions. The physical ``threshold'' begins at
$M_{J/\psi}^2\approx 9.6\,\mbox{GeV}^2$, but the perturbative
approximation is expected to fail earlier. Model-dependent 
studies of the charmonium contributions suggest that the 
perturbative approximation should be valid up to 
$q^2\approx (6-7)\,\mbox{GeV}^2$ \cite{Lim:1989yu,Kruger:1996cv}. 
(This supposes that we discard 
a charm mass as small as $1.2\,$GeV in estimating effects related 
to the charm threshold.)

\section{The decay $\bar{B}\to \bar K^*\gamma$}
\label{sec:gamma}

The decay rate for $\bar{B}\to \bar K^*\gamma$ in the heavy quark
limit is given by
\begin{equation}
\label{kgamrate}
\Gamma(\bar{B}\to \bar K^*\gamma) = \frac{G_F^2 |V_{ts}^*V_{tb}|^2}{8\pi^3}\,
M_B^3\left(1-\frac{m_{K^*}^2}{M_B^2}\right)^{\!3}
\frac{\alpha_{\rm em}}{4\pi}\,m_{b,\rm PS}^2\,\xi_\perp(0)^2
\left|{\cal C}_7\right|^2,
\end{equation}
with $m_{b,\rm PS}\equiv m_{b,\rm PS}(2\,\mbox{GeV})$ the PS mass. 
We included the kaon mass squared in a phase space and kinematic 
correction, but we generally neglect such terms in the dynamical quantity 
${\cal T}_\perp(0)=\xi_\perp(0)\,{\cal C}_7$. By helicity 
conservation the kaon is transversely polarized and only the 
transverse soft form factor $\xi_\perp(0)$ appears in the decay 
rate. In numerical form, the branching fraction is
\begin{equation}
\mbox{Br}(\bar{B}\to \bar K^*\gamma) = (7.9^{+1.8}_{-1.6})\cdot 10^{-5} 
\left(\frac{\tau_B}{1.6\mbox{ps}}\right) 
\left(\frac{m_{b,\rm PS}}{4.6\,\mbox{GeV}}\right)^{\!2}
\left(\frac{\xi_\perp(0)}{0.35}\right)^{\!2} 
= (7.9^{+3.5}_{-3.0})\cdot 10^{-5}
\end{equation}
using $\left|{\cal C}_7\right|^2 = 0.175^{+0.029}_{-0.026}$ 
and $m_{b,\rm PS}=(4.6\pm 0.1)\,$GeV, 
$\xi_\perp(0)=0.35\pm 0.07$, $|V_{ts}|=0.041\pm 0.003$, 
$\alpha_{\rm em}=1/137$, as in 
Table~\ref{tab2}. ($\tau_B$ denotes the $B$ meson lifetime.) 
For comparison 
we note that the leading order prediction is $4.5\cdot 10^{-5}$. The large 
difference reflects the enhancement of  $\left|{\cal C}_7\right|^2$ 
at next-to-leading order as discussed in Section \ref{sec:output}. 
The branching ratio predicted for the default parameter set is nearly 
twice as large as the current experimental averages
\cite{Coan:2000kh,babarmoriond,bellemoriond} 
\begin{eqnarray}
\label{datastar}
\mbox{Br}(\bar{B}^0\to \bar K^{*0}\gamma)_{\rm exp} &=& 
(4.54\pm 0.37)\cdot 10^{-5}, \\
\mbox{Br}(B^-\to \bar K^{*-}\gamma)_{\rm exp}&=& 
(3.81\pm 0.68)\cdot 10^{-5}. 
\end{eqnarray}
Note that we predict no difference of the neutral and charged $B$
meson decay rates at leading order in the heavy quark expansion, so
that the main effect of isospin breaking on branching fractions is 
the different lifetimes of the charged and neutral  mesons. 

Before speculating on the origin of this discrepancy, we note that 
there are alternative ways of representing the result of our
calculation. Since the decay amplitude is proportional to a single 
form factor, $T_1^{K^*}(0)$, there is no need to introduce the 
soft form factor $\xi_\perp(0)$, and we can express the decay rate 
directly in terms of the full QCD form factor. (The need to introduce 
the soft form factors arises when we consider the $\bar{K}^*\ell^+\ell^-$
final state, which involves many form factors that all relate to the 
same soft form factors.) Furthermore the dominant dependence on the 
$b$ quark mass arises through the factor $m_b$ in the definition of
the operator ${\cal O}_7$, so it also more natural to keep the 
$\overline{\rm MS}$ mass $\hat m_b(\hat m_b)$ as a prefactor rather
than the PS mass. This alternative 
representation amounts to a redefinition of ${\cal C}_7$ in which 
all {\em factorizable} corrections vanish at the scale $\hat m_b$, if 
we understand $T_1^{K^*}(0)$ as renormalized at the scale $\hat m_b$. We 
have computed the coefficient ${\cal C}^\prime_7$ also in this alternative 
scheme (indicated by the prime) and find that (\ref{kgamrate}) 
is modified as follows:
\begin{equation}
\label{schemes}
\begin{array}{ccc}
m_{b,\rm PS}^2\,\xi_\perp(0)^2\left|{\cal C}_7\right|^2
&\hspace*{0.3cm}\longrightarrow\hspace*{0.3cm}&
\hat m_b(\hat m_b)^2\,T_1^{K^*}(0)^2\left|{\cal C}^\prime_7\right|^2 
\\[0.3cm]
\left|{\cal C}_7\right|^2 = 0.175^{+0.029}_{-0.026}
&\hspace*{0.3cm}\longrightarrow\hspace*{0.3cm}&
\left|{\cal C}^\prime _7\right|^2 = 0.165^{+0.018}_{-0.017}
\end{array}
\end{equation}
The error is smaller in the primed scheme, because the hard-scattering
correction is much reduced once the factorizable correction is 
reabsorbed into the full QCD form factor. This implies less
sensitivity to the uncertain parameters that normalize this
correction. The two values given in (\ref{schemes}) 
are consistent with each other 
if one relates $T_1^{K^*}(0)=1.08\,\xi_\perp(0)$  \cite{Beneke:2001wa} 
and takes into 
account that $m_{b,\rm PS}(2\,\mbox{GeV})=4.6\,$GeV corresponds 
to $\hat m_b(\hat m_b)=4.4\,$GeV.

Several facts may explain the difference between the predicted and 
observed decay rate. The first and most interesting possibility is a 
modification of the standard model at short distances that would 
result in a smaller value of $|C_7|$. However, the modification needed 
to explain the observed decay rate is far too large not to be ruled out
by the agreement between experiment and theory for the 
{\em inclusive} decay $\bar B\to X_s\gamma$. It appears equally
implausible that new interactions would modify only the spectator
scattering and hence not show up in the inclusive rate. We therefore
conclude that the explanation must be sought in our understanding of
QCD effects. Most of the NLO enhancement is related to the
non-factorizable form factor type correction which appears in a 
nearly identical form in the inclusive decay rate as well. Hence 
this enhancement cannot be simply dismissed without putting the 
agreement for the inclusive decay into question. A second and 
less interesting possibility is therefore to invoke a large 
$1/m_b$ correction that would render our calculation unreliable. Given
the smallness of the  non-factorizable hard scattering correction 
it is not obvious how such a large dynamical enhancement of $1/m_b$ 
terms 
could be explained. This is in particular so as large $1/m_b$ 
corrections known to exist for non-leptonic decays 
\cite{Beneke:1999br,Beneke:2001ev} such as ``chirally enhanced''
corrections and large weak annihilation contributions are absent 
for decays into vector mesons. 

\begin{figure}[t]
\vspace*{3.6cm}
\hspace*{-0cm}
\epsffile{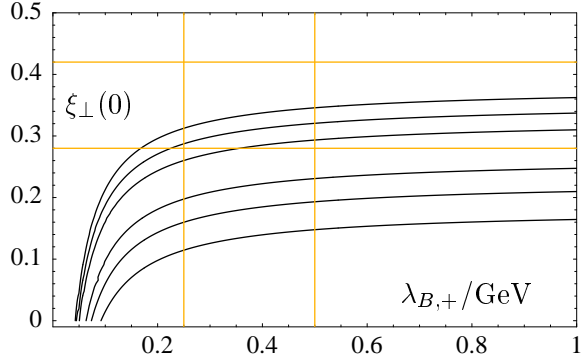}
\vskip0.1cm
\centerline{\parbox{14cm}{\caption{\label{fig:xiTfig}
Fit to $\xi_\perp(0)$ and $\lambda_{B,+}$. The solid curves give the 
$1\sigma$, $2\sigma$ and $3\sigma$ regions that follow from the 
observed $\bar B\to \bar K^*\gamma$ decay rate. The light solid curves 
mark the ranges assumed for $\xi_\perp(0)$ and $\lambda_{B,+}$ 
in Table~\ref{tab2}.}}}
\end{figure}

We therefore consider seriously the 
possibility that the form factors at $q^2=0$ are substantially 
different from what they are assumed to be in the QCD sum rule 
approach or in quark models. This possibility is entertained in 
Figure~\ref{fig:xiTfig}, where we show how the experimental 
data constrains the soft form factor $\xi_\perp(0)$ and the 
normalization of the hard scattering term. (We take all parameters 
other than $\lambda_{B,+}^{-1}$ 
in the normalizing factor constant and let $\lambda_{B,+}^{-1}$ 
be a representative. That is, a variation of  $\lambda_{B,+}^{-1}$ 
should be considered as a variation of a combination of parameters 
of the type (\ref{combpar}). All other parameter uncertainties 
are approximately accounted for by adding a $20\%$ uncertainty 
to the measured branching fraction which we combine with the
experimental error in quadrature.) The result of this fit is 
that 
\begin{equation}
T_1^{K^*}(0)_{|\mu=m_b}=0.27\pm 0.04 \qquad [\xi_\perp(0) = 0.24\pm 0.06],
\end{equation} 
where the error may be loosely 
interpreted as a one standard deviation error. This determination 
yields a smaller form factor than a similar determination 
in \cite{Burdman:2001ku}, where the 
NLO correction is not included and the bottom quark mass is handled 
differently.

Our results can be extended in a straightforward way to the 
decay $\bar B\to \rho\gamma$. This decay is particularly interesting in
searches for extensions of the standard model, because of the 
suppression of $b\to d$ transitions in the standard model and the 
simultaneous chirality suppression. Except for trivial adjustments 
the most important difference to the $\bar K^*\gamma$ final state 
is that the different flavour structure allows weak annihilation to 
proceed through the current-current operator with large Wilson
coefficient $\bar C_2$. This annihilation contribution is
power-suppressed in $1/m_b$, but the suppression is compensated 
by the large Wilson coefficient and the occurrence of annihilation 
at tree level. The annihilation contribution is calculable 
and has to be taken into account in a 
realistic analysis \cite{Grinstein:2000pc, Ali:2000zu}. 
This will be done elsewhere \cite{gerhard, us}, and we restrict
ourselves to $b\to s$ transitions in this paper.

\section{The decay $\bar{B}\to \bar K^* \ell^+ \ell^-$}
\label{sec:ll}

For the decay $\bar{B}\to \bar K^* \ell^+ \ell^-$ we obtain the double 
differential
decay spectrum (summed over final state polarisations,
and lepton masses neglected) 
\begin{eqnarray}
\frac{d^2\Gamma}{dq^2 d\!\cos\theta} 
&=& \frac{G_F^2 |V_{ts}^*V_{tb}|^2}
{128\pi^3}\,M_B^3\,\lambda(q^2,m_{K^*}^2)^3
\left(\frac{\alpha_{\rm em}}{4\pi}\right)^{\!2}\nonumber\\
&&
\Bigg[(1+\cos^2\theta)\,\frac{2 q^2}{M_B^2}\,\xi_\perp(q^2)^2 
\left(\,|{\cal C}_{9,\,\perp}(q^2)|^2+C_{10}^2\right)
\nonumber\\
&&
+\,(1-\cos^2\theta)\,
\left(\frac{E\,\xi_\parallel(q^2)}{m_{K^*}}\right)^{\!2}\,
\left(\,|{\cal C}_{9,\,\parallel}(q^2)|^2+C_{10}^2\,\Delta_\parallel(q^2)^2\right)
\nonumber\\
&&-\,\cos\theta\,\frac{8 q^2}{M_B^2}\,\xi_\perp(q^2)^2\,
\mbox{Re}({\cal C}_{9,\,\perp}(q^2)) \,C_{10}
\Bigg],
\label{dGamma}
\end{eqnarray}
where 
\begin{equation}
\lambda(q^2,m_{K^*}^2) = \Bigg[\left(1-\frac{q^2}{M_B^2}\right)^2-
    \frac{2 m_{K^*}^2}{M_B^2}
\left(1+\frac{q^2}{M_B^2}\right)+\frac{m_{K^*}^4}{M_B^4}\Bigg]^{\!1/2},
\end{equation}
and $q^2$ is the invariant mass of the lepton pair. The angle $\theta$
refers to the angle between the positively charged lepton and the 
$B$ meson in the center-of-mass frame of the lepton pair. We have 
kept terms of order $m_{K^*}^2$ in an overall factor related to 
phase space and kinematics but emphasize that the expression in 
brackets neglects such terms. The first two terms with angular 
dependence $(1\pm \cos^2\theta)$ correspond to the production of 
transversely and longitudinally polarized kaons, respectively. 
The third term generates a forward-backward asymmetry with respect to 
the plane perpendicular to the lepton momentum in the center-of-mass 
frame of the lepton pair. The factor $\Delta_\parallel$ multiplying 
the Wilson coefficient
$C_{10}$ in (\ref{dGamma}) arises from the factorizable
corrections to the form-factor $A_2$ and is
given by (cf.\/ Eq.~(66) in \cite{Beneke:2001wa})
\begin{eqnarray}
\Delta_\parallel(q^2) = 1+\frac{\alpha_s C_F}{4\pi}\,\big(-2+2 L\big) 
-\frac{\alpha_s C_F}{4\pi}\,\frac{2 q^2}{E^2}\,
\frac{\pi^2 f_B f_{K^*\,\parallel}\,\lambda_{B\,+}^{-1}}{N_c M_B (E/m_{K^*})
    \xi_\parallel(q^2)}\,\int_0^1\!\frac{du}{\bar u}\,\Phi_\parallel(u)
\end{eqnarray}
with $L$ defined in (\ref{Ldef}) and $E=(M_B^2-q^2)/(2\,M_B)$. 
(This factor could be eliminated 
by choosing a factorization convention different from (\ref{ins}) for  
the soft form factor $\xi_\parallel$ and by redefining 
${\cal C}_{9,\parallel}$ accordingly.)

Without going into much detail,
we remark that the result for the decay into pseudoscalar mesons,
$\bar B \to \bar K \ell^+\ell^-$, is easily obtained from the 
corresponding expressions for the decay into longitudinally polarized
vector mesons. The angular distribution is predicted to be proportional 
to $(1-\cos^2\theta)$ and for the lepton invariant mass spectrum we obtain
\begin{eqnarray}
\frac{d\Gamma}{dq^2} 
&=& \frac{G_F^2 |V_{ts}^*V_{tb}|^2}
{96\pi^3}\,M_B^3\,\lambda(q^2,m_K^2)^3 \left(\frac{\alpha_{\rm em}}{4\pi}
\right)^{\!2}
\xi_P(q^2)^2 
\left( |{\cal C}_{9,\,P}(q^2)|^2+C_{10}^2 \right)
\label{dGammaP}
\end{eqnarray}
where the quantities analogous to those in
(\ref{caltdef}), (\ref{CC9T}) are now defined as 
\begin{eqnarray}
 {\cal C}_{9,\,P}(q^2) &=& C_9 + \frac{2 m_b}{M_B} \, 
\frac{{\cal T}_P(q^2)}{\xi_P(q^2)},
\end{eqnarray}
 and 
\begin{eqnarray}
\langle \gamma^*(q,\mu) \bar K(p')| H_{\rm eff} | \bar{B}(p)
\rangle
 &=& \frac{G_F}{\sqrt{2}}\,V_{ts}^* V_{tb}\,\frac{g_{\rm em}
  m_b}{4\pi^2}
\,
\frac{{\cal T}_P(q^2)}{M_B}
\left[q^2 \, (p^\mu+p^{\prime\,\mu}) - (M_B^2-m_K^2) q^\mu \right].
\cr && 
\end{eqnarray}
The non-factorizable contributions to ${\cal T}_P$
are the same as those to ${\cal T}_\parallel$ up to an overall sign
and the replacement $\Xi_\parallel f_{K,\,\parallel}
\Phi_{K^*,\,\parallel}(u) \to f_K\Phi_{K}(u)$ in (\ref{nlodef}).
The factorizable contributions follow from the 
ratio of the tensor form factor $f_T(q^2)$ and the soft form factor
$\xi_P(q^2)$ as calculated in \cite{Beneke:2001wa}. 
The net result for ${\cal C}_{9,P}$ is then
\begin{equation}
\label{cP}
{\cal C}_{9,\,P}(q^2) = \frac{{\cal C}_{9,\,\parallel}(q^2)}
{\Delta_\parallel(q^2)},
\end{equation}
where the replacements $\Xi_\parallel f_{K,\,\parallel}
\Phi_{K^*,\,\parallel}(u) \to f_K\Phi_{K}(u)$, $\xi_\parallel(q^2)\to 
-\xi_P(q^2)$ are understood to be 
performed on the right-hand side.

\begin{figure}[t]
\vspace*{3.6cm}
\hspace*{-0cm}
\epsffile{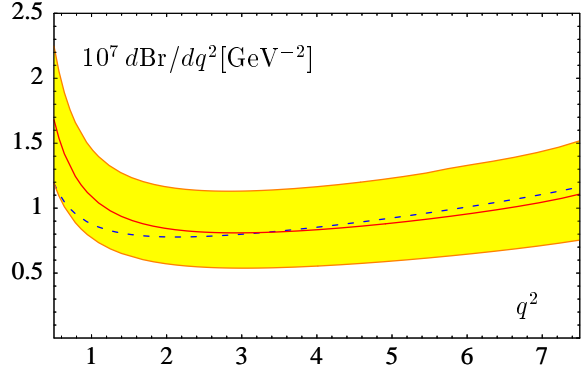}
\vskip0.1cm
\centerline{\parbox{14cm}{\caption{\label{fig:dgamma}
Differential decay rate $d\mbox{Br}(B^-\to {K^*}^-\ell^+\ell^-)/dq^2$ 
at next-to-leading order (solid center line) and leading order 
(dashed). The band reflects all theoretical uncertainties from 
parameters and scale dependence combined, with most of the 
uncertainty due to the form factors $\xi_{\perp,\,\parallel}(0)$. 
}}}
\end{figure}

\subsection{Lepton invariant mass spectrum}

The lepton invariant mass spectrum $d\mbox{Br}/d q^2$, obtained by 
integrating (\ref{dGamma}) over $\cos\theta$, is shown in
Figure~\ref{fig:dgamma}. (We use the $B$ meson lifetime 
$\tau_{B^-}=1.65\mbox{ps}$.) 
As the decay rate is dominated by the contribution
of longitudinally polarized $K^*$~mesons and the contribution 
from ${\cal O}_{10}$ except for small 
$q^2$, the impact of the 
next-to-leading order correction is small for $q^2>2\,\mbox{GeV}^2$. 
(This can be directly inferred from the small correction 
to $|{\cal C}_{9,\,\parallel}|$ as seen in Figure~\ref{fig:c9fig}.)
The apparently rather large 
uncertainty of our prediction is mainly due the form factors 
with their current large uncertainty and to a lesser extent due to 
$|V_{ts}|$ and the top quark mass. It may be hoped that in the 
longer term future the form factors could be known with much greater 
confidence. The uncertainties can then be reduced by a
factor of three in which case the limiting factor of our prediction 
is probably $1/m_b$ corrections, all other parameter and scale 
variations being negligible in comparison. 

Since $|{\cal C}_{9,\,\parallel}|$ depends on the charge of the decaying
$B$ meson, there is a small amount of isospin breaking. While this 
effect is $q^2$-dependent for $|{\cal C}_{9,\parallel}|$, the 
combination of coefficients that enters the lepton invariant mass 
spectrum turns out to be almost independent on $q^2$.  The effect is, 
however, very small,
\begin{equation}
\delta = \frac{d\Gamma(B^-\to K^{*-}\ell^+\ell^-)/dq^2-
d\Gamma(\bar{B}^0\to \bar{K}^{*0}\ell^+\ell^-)/dq^2}
{d\Gamma(B^-\to K^{*-}\ell^+\ell^-)/dq^2+
d\Gamma(\bar{B}^0\to \bar{K}^{*0}\ell^+\ell^-)/dq^2} 
\approx  1\%.
\end{equation}
For the differential branching fractions the main isospin breaking
effect arises from the lifetime difference of the charged and neutral 
$B$ mesons.

The impact of the NLO correction on the lepton invariant mass 
spectrum to the decay in a pseudoscalar meson follows a qualitatively 
similar pattern as for the vector meson final state, in particular 
as the decay rate involves only a quantity closely related to 
$|{\cal C}_{9,\,\parallel}|$ (see (\ref{cP})). We therefore
do not discuss this decay in further detail here, but note that the 
lepton invariant mass spectrum develops a logarithmic singularity 
for small $q^2$. This is due to the long-distance sensitivity
mentioned above, which is now dominant because the photon pole is 
absent. The invariant mass spectrum is therefore non-perturbative 
for $q^2\sim \Lambda_{\rm QCD}^2$, but perturbative for 
$q^2\sim m_b\Lambda_{\rm QCD}$. Furthermore, the non-perturbative 
contribution is formally power-suppressed when the lepton invariant
mass spectrum is integrated from 0 to some 
$q^2$ of order $m_b\Lambda_{\rm QCD}$.

\subsection{Forward-backward asymmetry}
\label{sec:AFB}

The QCD factorization approach proposed here 
leads to an almost model-independent theoretical prediction for 
the forward-backward asymmetry \cite{Ali:1991is}. 
It has been noted in \cite{Burdman:1998mk} that the 
location of the forward-backward asymmetry zero is nearly 
independent of particular form factor models. An explanation 
of this fact was given in \cite{Ali:1999mm}, 
where it has been noted that the form factor ratios on which the asymmetry 
zero depends are predicted free of hadronic uncertainties in 
the combined heavy quark and large energy limit. 
In \cite{Beneke:2001wa} the effect of 
the (factorizable) radiative corrections to the form factor ratios
has been studied and has been found to shift the position of the asymmetry
zero about 5\% towards larger values. 
We are now in the position to discuss 
the effect of both, factorizable {\em and}\/ non-factorizable 
radiative corrections to next-to-leading order in the strong
coupling constant on the location of the asymmetry-zero, and hence 
to complete our earlier analysis.

\begin{figure}[t]
\vspace*{3.6cm}
\hspace*{-0cm}
\epsffile{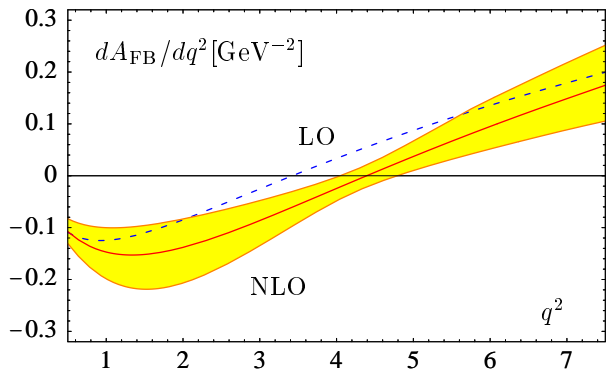}
\vskip0.3cm
\centerline{\parbox{14cm}{\caption{\label{fig:dafbnorm}
Forward-backward asymmetry 
$d A_{\rm FB}(B^-\to K^{*-}\ell^+\ell^-)/dq^2$ 
at next-to-leading order (solid center line) and leading order 
(dashed). The band reflects all theoretical uncertainties from 
parameters and scale dependence combined. }}}
\end{figure}

We define the forward-backward (FB) asymmetry
(normalized to the differential decay rate $d\Gamma(B^- \to 
K^*{}^- \ell^+\ell^-)/dq^2$) by 
\begin{eqnarray}
 \frac{dA_{\rm FB}}{dq^2} &\equiv &
 \frac{1}{d\Gamma/dq^2} \left(\,
\int_0^1 d(\cos\theta) \,\frac{d^2\Gamma}{dq^2 
d\cos\theta} - \int_{-1}^0 d(\cos\theta) \,\frac{d^2\Gamma}{dq^2 
d\cos\theta} \right) 
\end{eqnarray}
Our result for the FB asymmetry is shown in Figure~\ref{fig:dafbnorm}
to LO and NLO accuracy.
From (\ref{dGamma}) it is obvious that
$dA_{\rm FB}/dq^2
\propto\mbox{Re}\left({\cal C}_{9,\perp}(q^2)\right)$,
and therefore the FB asymmetry vanishes if
$\mbox{Re}\left({\cal C}_{9,\perp}(q_0^2)\right)=0$.
At leading order this translates into the relation
\begin{eqnarray}
 C_9 +  \mbox{Re}(Y(q_0^2))
 &=&  - \frac{2 M_B m_b}{q_0^2} \, C_7^{\rm eff} \ ,
\end{eqnarray}
which, as already mentioned, is free of hadronic uncertainties.
This effect can also be seen in Figure~\ref{fig:dafbnorm} where
the uncertainty for the prediction of the
FB asymmetry gets smaller around $q_0^2$.
Using our default input parameters we obtain 
$q_0^2=3.4^{+0.6}_{-0.5}$~GeV$^2$ at LO where the error is
by far dominated by the scale dependence.
The numerical effect of NLO corrections
amounts to a substantial reduction of the FB asymmetry for  
intermediate lepton invariant mass 
($q^2= 2-5$~GeV$^2$) 
and a significant shift of the location of the asymmetry zero to
\begin{eqnarray}
  q_0^2 &=& 4.39 {}^{+ 0.38}_{-0.35} ~{\rm GeV}^2
\label{q0estimate}
\end{eqnarray}
The largest single uncertainty (about $\pm 0.25\,\mbox{GeV}^2$) 
continues to be 
scale dependence (varying $\mu$ within 
$m_b/2 < \mu < 2 m_b$ as usual), but also the
variation of $m_b$, $\Lambda_{\rm QCD}$, and $\lambda_{B,+}$
gives an effect.

Our LO result for $q_0^2$ differs somewhat from the value 
$q_0^2 = 2.88 \, {}^{+0.44}_{-0.28}$ quoted 
in \cite{Ali:1999mm}.
The reason for this difference is a combination of three effects:
(i) slightly different values for the SM Wilson coefficients, 
in particular a larger value for $C_9$ in \cite{Ali:1999mm}, 
(ii) a different treatment of
the $b$-quark mass, (iii) the inclusion of (higher-order factorizable)
form factor corrections in \cite{Ali:1999mm}, since the full QCD 
form factors estimated from light-cone 
QCD sum rules \cite{Ball:1998kk} are used. The first two issues 
are resolved within the present next-to-leading order treatment 
but with respect to (iii) we  
note that the form factor ratios entering ${\cal C}_{9,\perp}$
are exactly those where there exists a discrepancy 
between the QCD sum rule result and the heavy quark limit 
\cite{Beneke:2001wa}.
The origin of this discrepancy has not been clarified, 
but it is conceivable that it is related to some $1/M_B$
corrections, which are implicitly included in the sum rule estimate.
In order to estimate the impact of using full QCD form factors on 
the location of the asymmetry zero, we reabsorb the factorizable 
$\alpha_s$ corrections
into the physical form factors and the $\overline{\rm MS}$ mass. 
The FB asymmetry then reads (with the tensor form factors and 
the $\overline{\rm MS}$ mass $\hat{m}_b$ all renormalized at the scale 
$\mu$)
\begin{eqnarray}
\label{fbnew}
\frac{dA_{\rm FB}}{dq^2} &=& -\frac{1}{d\Gamma/dq^2} \,
\frac{G_F^2 |V_{ts}^*V_{tb}|^2}
{128\pi^3}\,M_B^3\,\lambda(q^2,m_{K^*}^2)^2
\left(\frac{\alpha_{\rm em}}{4\pi}\right)^{\!2}\, 
\frac{8 q^2}{M_B^2} \, C_{10} \,A_1(q^2) V(q^2)
\nonumber\\[0.2em]
&&\times \mbox{Re} \Bigg[\left( C_9 +  Y(q^2) 
       + \frac{\alpha_s C_F}{4\pi} \,C_\perp^{(\rm nf,9)}(q^2)\right)
\nonumber \\[0.2em]
&& \qquad+\,\frac{\hat m_b}{q^2}\,
\left((M_B + m_{K^*})\,\frac{T_1(q^2)}{V(q^2)}+
(M_B - m_{K^*}) \,\frac{T_2(q^2)}{A_1(q^2)} \right)
\nonumber \\[0.2em] 
&& \qquad\qquad \times \left(C_7^{\rm eff}  
       + \frac{\alpha_s C_F}{4\pi} \, C_\perp^{(\rm nf,7)}(q^2)
\right)\nonumber\\
&&  \qquad+\,\frac{\hat m_b}{q^2}\,\left((M_B + m_{K^*})\,\frac{1}{V(q^2)}+
(M_B - m_{K^*})\,\left(1-\frac{q^2}{M_B^2}\right) \,\frac{1}{A_1(q^2)} 
\right) \nonumber\\
&&  \qquad\qquad \times \, \frac{\alpha_s C_F}{4\pi} \, \frac{\pi^2}{N_c}
\,\frac{f_B f_{K^*,\,\perp} \, \lambda_{B,+}^{-1}}{M_B} \,
\int_0^1 \!du\,\Phi_{K^*,\,\perp}(u) \,T_{\perp,\,+}^{({\rm
    nf})}(u)
\Bigg]
\end{eqnarray}
The non-factorizable contributions from  
$C_\perp^{(\rm nf)}$ in (\ref{start3}) are now divided into two
parts, coming with the operator structure of ${\cal O}_7$ and 
${\cal O}_9$, respectively,
\begin{eqnarray}
C_F C^{(\rm nf,9)}_{\perp}(q^2) &=& 
- \left[\bar{C}_2 F_2^{(9)}(q^2)+2\bar{C}_1\left(F_1^{(9)}(q^2)+
\frac{1}{6}\,F_2^{(9)}(q^2)\right)+C_8^{\rm eff} F_8^{(9)}(q^2)\right] , \\
C_F C^{(\rm nf,7)}_{\perp}(q^2) &=& - \left[\bar{C}_2 F_2^{(7)}(q^2)+
C_8^{\rm eff} F_8^{(7)}(q^2)\right].
\end{eqnarray}
These are just the terms that also enter the inclusive decay, while the 
last term in (\ref{fbnew}) arises from the hard spectator-scattering and thus
only appears in the exclusive decay.
Using the relations $T_2 = 2 E T_1/M_B$ and $A_1 = 2 E M_B V/(M_B+m_{K^*})^2$, 
which are presumed to hold to all orders in $\alpha_s$, and to leading 
order in $1/m_b$ \cite{Beneke:2001wa,Burdman:2001ku}, (\ref{fbnew}) 
can be rearranged to contain only $V$ and $T_1/V$. If furthermore 
the latter ratio is expanded in $\alpha_s$ and if some terms of order 
$m_{K^*}^2/M_B^2$ are neglected, we return to the next-to-leading order 
result that we use by default.
On the basis of (\ref{fbnew}) and the form factor estimates 
from \cite{Ball:1998kk}, we now obtain 
$q_0^2 = 3.25$~GeV$^2$ at LO and $q_0^2=3.94$~GeV$^2$ at NLO. 
(To obtain this result we use $m_{b,\rm PS}$ as input and 
treat perturbatively the difference with $\hat m_b$.) 
The NLO value is slightly more than  
one ``standard deviation'' away from our default result in
(\ref{q0estimate}). The difference between the two values 
could be viewed as an estimate for an additional systematic error, 
but the main conclusion is that using the soft form factors or the 
full QCD form factors from light-cone QCD sum rules 
does not cause a large ambiguity in our result for $q_0^2$. We 
therefore assume
\begin{eqnarray}
\label{q0result2}
  q_0^2 &=& (4.2\pm 0.6)\,{\rm GeV}^2
\end{eqnarray}
to be a conservative estimate for the asymmetry zero.

\begin{figure}[t]
\vspace*{3.6cm}
\hspace*{-0cm}
\epsffile{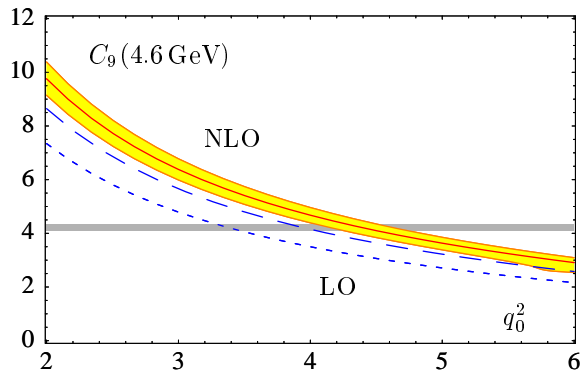}
\vskip0.3cm
\centerline{\parbox{14cm}{\caption{\label{fig:c9exp}
The Wilson coefficient $C_9(m_b)$ deduced from a measurement of 
the asymmetry zero $q_0$ at next-to-leading order including all 
parameter uncertainties (light band), but not including the slightly 
enlarged error in (\ref{q0result2}). For comparison we also show 
(without uncertainties) the leading order result (dashed) and 
the result at next-to-leading order using (\ref{fbnew}) and
the full QCD form factors from \cite{Ball:1998kk} (long-dashed). 
The horizontal band is the standard model value $C_9(4.6\,\mbox{GeV})=
4.21\pm 0.12$.}}}
\end{figure}

The Wilson coefficients may also receive
contributions from new particles in theories beyond the standard model.
(Of course extensions of the standard model may introduce a 
larger set of operators as well.)
The function ${\cal C}_{9,\perp}$ depends dominantly on
$C_9$ and $C_7^{\rm eff}$ (see (\ref{CC9T})).
The experimental measurement of the decay rate for
the inclusive decay $\bar B \to X_s \gamma$ already constrains 
$|C_7^{\rm eff}|$ to be close to its standard model value. 
The forward-backward asymmetry will establish the sign of 
$C_7^{\rm eff}$ and assuming this to take its standard model 
value, a measurement of the FB asymmetry zero in 
the decay $\bar B \to \bar K^* \ell^+ \ell^-$ provides a way to determine 
$C_9$. 
It is therefore instructive to consider our result as a prediction 
for $C_9$ for a given value of $q_0^2$ that will be
measured in future experiments. 
In Figure~\ref{fig:c9exp} we show our NLO prediction for
$C_9$ at the scale
$\mu = 4.6$~GeV as a function of $q_0^2$ together with the
standard model 
prediction (see Section~\ref{input_sec}). For comparison we also
show the LO result and the modification arising from using
(\ref{fbnew}) instead of (\ref{dGamma}). We remark that $C_9$ is 
scheme-dependent and so the plot refers to the particular 
renormalization scheme for ${\cal O}_9$ adopted in \cite{Chetyrkin:1997vx}.

Figure~\ref{fig:c9exp} is our main phenomenological result. It  
illustrates that the FB asymmetry zero can be used to test the
standard model and to search for extensions, since the theoretical
uncertainties relevant for the determination of $C_9$ are under
control. In this respect the systematic inclusion of NLO corrections,
which is provided in this work, turns out to be essential, because
it reduces the renormalization scale and scheme dependences to
a large extent and corrects the leading-order result by a large amount.
We emphasize that in contrast to the absolute rates for $\bar B \to
\bar K^*\gamma$ and  $\bar B \to \bar K^*\ell^+\ell^-$ the 
dependence on the $\bar B \to \bar K^*$ form factors is 
sub-leading for the FB asymmetry zero. 
Thus the current difficulty to account for the 
$\bar B \to
\bar K^*\gamma$ branching ratio (which may suggest a smaller value 
of the form factor
$T_1(0)$ than usually estimated from QCD sum rules as discussed above), 
does not necessarily cause a problem for our prediction of 
$C_9$ vs.\ $q_0^2$. We also note that the resulting values of 
$q_0^2$ are sufficiently below the threshold for charmonium
resonances ($J/\psi$, $\psi'$), so that a long-distance 
contribution to the function $Y(q^2)$ can safely be neglected. 
We therefore conclude that if the asymmetry zero is found 
at some $q_0^2<6\,\mbox{GeV}^2$, the short-distance coefficient 
$C_9(4.6\,\mbox{GeV})$ can be determined with a theoretical 
error of $\pm 10\%$. If its value turns out to be different from 
the standard model value $4.21\pm 0.12$ by an amount $\Delta$ 
and if we assume that 
all Wilson coefficients other than $C_{9,10}$ remain unmodified 
at the scale $M_W$, then (\ref{rgesol2}) below implies a new contribution 
of magnitude $\Delta$ to the coefficient function $C_9$ at the 
scale $M_W$.

\section{Concluding discussion}
\label{sec:final}

Building on our previous work on heavy-to-light form factors 
\cite{Beneke:2001wa} and on work on 
non-leptonic decays \cite{Beneke:1999br,Beneke:2000ry},
we demonstrated that also the radiative $B$ decays 
$B\to V\gamma$ and $B\to V\ell^+\ell^-$ can be computed in the 
heavy-quark limit. This allows us to solve the problem of 
non-factorizable strong interaction corrections (i.e.\ those
corrections not related to form factors), which has so far prevented 
a systematic discussion of exclusive radiative decays, as compared 
to their inclusive analogues. The approach presented here does 
not circumvent the need to know the heavy-to-light form factors, 
but we may hope that progress in lattice QCD will give us these 
form factors at smaller $q^2$ and more reliably than at present 
in the longer term future.

In the present paper we have concentrated on the $b\to s$ transition. 
We noted that the next-to-leading order correction yields an 
$80\%$ enhancement of the $B\to K^*\gamma$ decay rate, a large part 
of which is related to the NLO correction that appears also in the 
inclusive decay \cite{Greub:1996tg,Asatrian:2001de}. The enhancement 
is so large that the predicted decay rate disagrees with the 
observed decay rate unless the form factors at $q^2=0$ 
are much smaller than what they are believed to be, or unless our 
theoretical prediction is made unreliable by a large correction 
to the heavy quark limit. Perhaps the most interesting outcome 
of our analysis is that while the NLO correction to 
the lepton invariant mass spectrum in $B\to K^*\ell^+\ell^-$ is 
very small, there is a large correction to the predicted location 
of the forward-backward asymmetry zero. The particular interest 
in this quantity derives from the fact that all dependence of form 
factors arises first at next-to-leading order. This implies that 
(given the Wilson coefficient $C_7$) the Wilson coefficient $C_9$ 
can be determined from a measurement of the 
location of the asymmetry zero with little theoretical uncertainty. 
We find that this conclusion holds true in particular after 
including the NLO correction, which however, shifts the predicted 
location of the zero by $1\,\mbox{GeV}^2$, or about $30\%$, with 
a residual uncertainty estimated to be about 
$(0.4-0.6)\,\mbox{GeV}^2$. This implies that an accurate 
measurement of the asymmetry zero determines the Wilson coefficient 
$C_9$ at the scale $m_b$ with an error of $\pm 10\%$.

The method developed here can also be applied to $b\to d$ 
transitions. These are particularly interesting in the search for 
modifications of the standard model. The better control of 
standard model effects after accounting for non-factorizable 
corrections should help increasing the sensitivity to such
non-standard effects. In particular we note that the approach
presented here allows us to compute isospin breaking effects, which 
may turn out to be a particularly nice signal of non-standard
physics. A detailed discussion of $b\to d$ transitions requires,
however, a more careful study of weak annihilation effects, and we 
plan to return to non-standard physics in the context of such a 
study \cite{us}.

A systematic, but rather difficult to quantify, limitation of the 
factorization approach is the poor control of $1/m_b$ suppressed 
effects. These arise from various sources and we conclude by
mentioning one of them. We may ask to what extent a non-leptonic 
decay $B\to K^*\rho$ in which the $\rho$ meson is assumed to convert
to a photon through vector meson dominance may affect our 
calculation of the direct decay into $K^*\gamma^*$. More accurately, 
the question is to what extent the hadronic structure of the photon 
matters at small $q^2$. It is not difficult to see that the indirect 
contribution is suppressed by one power of $m_b$ in the heavy quark 
limit. For the sake of the argument we return to the vector dominance 
picture and note that there is a one-to-one correspondence 
between diagrams that appear in the factorization approach to 
non-leptonic decays \cite{Beneke:1999br,Beneke:2000ry} and 
the diagrams computed here. For example, the naive factorization 
contribution to the former case corresponds to our leading order weak 
annihilation term; the spectator scattering contribution to
non-leptonic decays corresponds to the diagram in Figure~\ref{fig1}b 
with the photon attached to the internal quark loop and so on. 
Counting powers of $m_b$ for all the quantities that appear in these 
amplitudes we find that the direct amplitude is always larger 
by a factor of $m_b$. 

\subsection*{Acknowledgements}

We would like to thank C.~Greub and M.~Walker 
for helpful discussions. 
While this work has been done, we became aware of related work 
in \cite{gerhard}. We thank G.~Buchalla for discussing the results 
of this paper prior to publication.

\subsection*{Note added}

While completing this work, the article~\cite{Ali:2001ez} appeared, 
in which the next-to-leading order hard-scattering 
correction to $B \to \rho \gamma$
is calculated. The corresponding quantity should be obtained
from the function $T_{\perp,+}^{(\rm nf)}$ in (\ref{Tnfperp})
in the limit $q^2 \to 0$. The small QCD penguin contributions 
have been neglected in \cite{Ali:2001ez}, so that only the terms 
proportional to $\bar{C}_2$ and $C_8^{\rm eff}$ have to be retained 
for the comparison. The charm quark mass in  $t_\perp(u,m_c)$ is 
set to 0 in \cite{Ali:2001ez}, presumably because an endpoint 
divergence is found for $m_c\not= 0$. We note that our result 
does not exhibit this endpoint divergence (which would invalidate 
the factorization approach), but setting $m_c=0$ for the sake 
of comparison, we obtain 
\begin{displaymath}
T_{\perp,+}^{(\rm nf)}(u,\omega) \to -\frac{4 e_d\,C_8^{\rm eff}}{u}+
\frac{M_B}{2 m_b}\,\frac{4 e_u \bar C_2}{\bar u}.
\end{displaymath}
We then differ from (4.12) of \cite{Ali:2001ez} by a sign in the first term 
and a factor of $-4$ in the second term. 
We also note that \cite{Ali:2001ez} 
includes a factorizable form factor correction from  \cite{Beneke:2001wa} 
while retaining the full QCD form factor $T_1(0)$. This implies that 
the factorizable correction is included twice. 

\begin{appendix}

\section{Operator bases}
\label{app:a}

We use the renormalization conventions of
\cite{Chetyrkin:1997vx}, but present our result in terms of the 
following linear combinations of Wilson coefficients:
\begin{eqnarray}
\bar{C}_1 &=& \frac{1}{2} \,C_1,\nonumber\\
\bar{C}_2 &=& C_2-\frac{1}{6}\,C_1,\nonumber\\
\bar{C}_3 &=& C_3-\frac{1}{6}\,C_4+16\,C_5-\frac{8}{3}\,C_6,\nonumber\\
\bar{C}_4 &=& \frac{1}{2}\,C_4+8\,C_6,\nonumber\\
\bar{C}_5 &=& C_3-\frac{1}{6}\,C_4+4\,C_5-\frac{2}{3}\,C_6,\nonumber\\
\bar{C}_6 &=& \frac{1}{2}\,C_4+2\,C_6.
\end{eqnarray}
These definitions hold to all orders in perturbation theory. The 
``barred'' coefficients are related to those 
defined in \cite{Buchalla:1996vs} by \cite{Chetyrkin:1998gb}
\begin{equation}
\bar{C}_i = C^{\rm BBL}_i +\frac{\alpha_s}{4\pi}\,T_{ij}\, C^{\rm
  BBL}_j + O(\alpha_s^2),
\end{equation}
where 
\begin{equation}
T = \left(\begin{array}{cccccc}
\frac{7}{3} & 2 & 0 & 0 & 0 & 0 \\[0.1cm]
1 & -\frac{2}{3} & 0 & 0 & 0 & 0 \\[0.1cm]
0 & 0 & -\frac{178}{27} & -\frac{4}{9} & \frac{160}{27} & \frac{13}{9}
\\[0.1cm]
0 & 0 & \frac{34}{9} & \frac{20}{3} & -\frac{16}{9} & -\frac{13}{3}
\\[0.1cm]
0 & 0 & \frac{164}{27} & \frac{23}{9} & -\frac{146}{27} & -\frac{32}{9}
\\[0.1cm]
0 & 0 & -\frac{20}{9} & -\frac{23}{3} & \frac{2}{9} & \frac{16}{3}
\end{array}\right).
\end{equation}

\section{Unexpanded form of $F_{8}^{(7,9)}$}
\label{app:b}

Using the conventions of  \cite{Asatrian:2001de} we obtain for 
the 1-loop matrix element of ${\cal O}_8$ (Figure~\ref{fig1}c together 
with a set of symmetric diagrams):

\begin{eqnarray}
  F_8^{(7)} &=& -\frac{32}{9} \, \ln \frac{\mu}{m_b} 
  - \frac 89 \, \frac{\hat s}{1- \hat s} \, \ln \hat s- \frac89 \, i\pi 
    - \frac 49 \, \frac{ 11 - 16 \hat s + 8 \hat s^2 }{(1-\hat s)^2}
\cr && \quad +
    \frac 49 \, \frac{1}{(1-\hat s)^3} \,
   \left[ 
     (9 \hat s - 5 \hat s^2 + 2 \hat s^3) \, B_0(\hat s)
    - (4 + 2 \hat s) \, C_0(\hat s) \right],
\\[0.3em]
  F_8^{(9)} &=& 
    \frac {16}{9} \, \frac{1}{1- \hat s}\, \ln \hat s
             + \frac{8}{9}  \, \frac{5 - 2 \hat s}{(1-\hat s)^2}
 -
    \frac 89 \, \frac{4 - \hat s}{(1-\hat s)^3} 
   \left[  (1 + \hat s) \, B_0(\hat s)
         - 2 \, C_0(\hat s)
   \right],       
\end{eqnarray}
where $\hat s = q^2/m_b^2$, $B_0(\hat s)=B_0(q^2,m_b^2)$ is given 
in (\ref{b0def}), and the integral 
\begin{eqnarray}
 C_0(\hat s) &=& \int_0^1 dx \, \frac{1}{x \, (1-\hat s) + 1} \, 
                 \ln \frac{x^2}{1- x \, (1-x) \, \hat s}
\end{eqnarray}
can be expressed in terms of dilogarithms.

\section{NNLL formulae for $C_9$ and $C_{1,...,6}$}
\label{app:c}

Because the anomalous dimension of ${\cal O}_9$ begins at order 
$\alpha_s^0$, the Wilson coefficient $C_9$ is needed to 
next-to-next-to-leading logarithmic (NNLL) accuracy. This requires 
also the coefficient of the four-quark operators to this accuracy. 
Although not all of the necessary inputs are currently known, we 
present here the solution of the renormalization group equations to 
NNLL order. 

We first restrict ourselves to the sector of four-quark operators. 
Define $a_s(\mu)=\alpha_s(\mu)/(4\pi)$ and 
$\beta(a_s)=\beta_0 a_s^2+\ldots$ with $\beta_0=11-2 n_f/3$. We
further define the matrix $U(\mu,M_W)$ such that
\begin{equation}
C(\mu) = U(\mu,M_W) \,C(M_W),
\end{equation}
where $C(\mu)$ is the vector of Wilson coefficients. $U(\mu,M_W)$ 
satisfies 
\begin{equation}
\label{rge}
\frac{d}{d\ln\mu} \,U(\mu,M_W) = \gamma^T(\mu) \, U(\mu,M_W),
\end{equation}
and the anomalous dimension matrix is expanded as 
\begin{equation}
\gamma = \gamma^{(0)} a_s+\gamma^{(1)} a_s^2+\ldots.
\end{equation}
Let $V$ be the matrix that diagonalizes ${\gamma^{(0)}}^T$, so that 
\begin{equation}
V^{-1} {\gamma^{(0)}}^T V =\left[\gamma_i^{(0)}\right]_{\rm diag},
\end{equation}
and define
\begin{equation}
U^{(0)}(\mu,M_W) = V\left[\left(\frac{a_s(\mu)}{a_s(M_w)}\right)^
{-\gamma_i^{(0)}/(2\beta_0)}\right]_{\rm diag} V^{-1},
\end{equation}
which solves (\ref{rge}) to leading order in $a_s(\mu)$. Then 
the NNLL expression for the evolution matrix $U(\mu,M_W)$ 
reads
\begin{eqnarray}
\label{rgesol1}
U(\mu,M_W)&=& \left(1+a_s(\mu) J^{(1)} + a_s(\mu)^2 J^{(2)}\right) 
 U^{(0)}(\mu,M_W) 
\nonumber\\
&&\hspace*{0.5cm}
\times 
\left(1-a_s(M_W) J^{(1)} - a_s(M_W)^2 \left[J^{(2)}-{J^{(1)}}^2\right]
\right),
\end{eqnarray}
where 
\begin{equation}
J^{(n)}= V H^{(n)} V^{-1}.
\end{equation}
The matrices $H^{(1)}$, $H^{(2)}$ have the entries 
\begin{eqnarray}
H^{(1)}_{ij} &=& \delta_{ij}\gamma^{(0)}_i\frac{\beta_1}{2\beta_0^2} 
-\frac{G^{(1)}_{ij}}{2\beta_0+\gamma^{(0)}_i-\gamma^{(0)}_j},
\\
H^{(2)}_{ij} &=& \delta_{ij}\gamma^{(0)}_i\frac{\beta_2}{4\beta_0^2} 
+\sum_k \frac{2\beta_0+\gamma^{(0)}_i-\gamma^{(0)}_k}
{4\beta_0+\gamma^{(0)}_i-\gamma^{(0)}_j} \left(H^{(1)}_{ik}H^{(1)}_{kj}-
\frac{\beta_1}{\beta_0} H^{(1)}_{ij}\delta_{jk}\right)
\nonumber\\
&&-\,\frac{G^{(2)}_{ij}}{4\beta_0+\gamma^{(0)}_i-\gamma^{(0)}_j}
\end{eqnarray}
and we have defined
\begin{equation}
G^{(n)}= V^{-1}  {\gamma^{(n)}}^T V.
\end{equation}
The same equations can also be used to solve the renormalization group
equations when the electromagnetic and chromomagnetic dipole operators are 
included. Then $\gamma$ is an $8\times 8$ matrix, and one needs 
to keep in mind that the $6\times 2$ block in $\gamma^{(n)}$ 
that mixes the dipole and four-quark operators is obtained from 
$n+2$ loop diagrams, while all other entries come from $n+1$ loop 
diagrams. In our numerical analysis we expand $U(\mu,M_W) \,C(M_W)$ 
in powers of $a_s(\mu)$ and $a_s(M_W)$ using (\ref{rgesol1}) and 
the expansion of the initial condition $C(M_W)$ in powers of 
$a_s(M_W)$ and keep only terms to the required order in both expansion
parameters. 

The renormalization group equation for $C_9$ is particularly simple
and can be obtained directly by integrating the coefficients of
four-quark operators. The equation reads
\begin{equation}
\label{rge2}
\frac{d}{d\ln\mu}\,C_9(\mu) = \kappa(\mu)\, C(\mu),
\end{equation}
where $C(\mu)$ is the vector of Wilson coefficients of four-quark
operators and 
\begin{equation}
\kappa = \kappa^{(-1)} +\kappa^{(0)} a_s + \ldots
\end{equation}
is the $1\times 6$ matrix that describes the mixing into 
${\cal O}_9$. The calculation of $\kappa^{(n)}$ involves $n+2$ loop 
diagrams. The solution of (\ref{rge2}) is given by
\begin{equation}
\label{rgesol2}
C_9(\mu) = C_9(M_W) + W(\mu,M_W)\,C(M_W),
\end{equation}
with the $1\times 6$ matrix
\begin{equation}
 W(\mu,M_W) = -\frac{1}{2}\,\int_{a_s(M_W)}^{a_s(\mu)}\!da_s 
\,\frac{\kappa(a_s)}{\beta(a_s)}\,U(\mu,M_W)
\end{equation}
with the evolution matrix $U(\mu,M_W)$ from the four-quark sector as 
defined above. The solution to NNLL accuracy is obtained by 
inserting $U(\mu,M_W)$ to this accuracy. We introduce the $6\times 6$
matrices 
\begin{equation}
D_{n}(\mu,M_W) = V\left[\frac{1}{n-\gamma_i^{(0)}/(2\beta_0)} 
\Bigg[\left(\frac{a_s(\mu)}{a_s(M_w)}\right)^
{\!n-\gamma_i^{(0)}/(2\beta_0)}-1\Bigg] \right]_{\rm diag} V^{-1}, 
\end{equation}
in terms of which the solution is given by
\begin{eqnarray}
\label{Wsol}
W(\mu,M_W) &=& -\frac{\kappa^{(-1)}}{2\beta_0 a_s(M_W)}\,
D_{-1}(\mu,M_W) 
\nonumber\\
&&\hspace*{-2cm}
-\,\frac{1}{2\beta_0} \Bigg[\left(\kappa^{(0)} - \frac{\beta_1}{\beta_0}\,
\kappa^{(-1)} + \kappa^{(-1)}\,J^{(1)}\right) D_{0}(\mu,M_W)
- \kappa^{(-1)}\,D_{-1}(\mu,M_W)\,J^{(1)}\Bigg]
\nonumber\\[0.2cm]
&&\hspace*{-2cm}
-\,\frac{a_s(M_W)}{2\beta_0} 
\Bigg[\Bigg(\kappa^{(1)} - \frac{\beta_1}{\beta_0}\,
\kappa^{(0)} +
\bigg(\frac{\beta_1^2}{\beta_0^2}-\frac{\beta_2}{\beta_0}
\bigg)\,\kappa^{(-1)}
+\bigg(\kappa^{(0)} - \frac{\beta_1}{\beta_0}\,\kappa^{(-1)}\bigg) 
J^{(1)}
\nonumber\\
&&\hspace*{-0.5cm}
+\,\kappa^{(-1)}\,J^{(2)}\Bigg) D_{1}(\mu,M_W)
- \left(\kappa^{(0)} - \frac{\beta_1}{\beta_0}\,
\kappa^{(-1)} + \kappa^{(-1)}\,J^{(1)}\right) D_{0}(\mu,M_W)\,J^{(1)}
\nonumber\\
&&\hspace*{-0.5cm}
-\, \kappa^{(-1)}\,D_{-1}(\mu,M_W)\,\Big[J^{(2)}-{J^{(1)}}^2\Big]\Bigg]
\end{eqnarray}
In our numerical analysis we expand (\ref{rgesol2})
in powers of $a_s(M_W)$ and keep only terms to the required order 
in the expansion parameter. To NNLL we thus need 
$C_9(M_W)$ to order $a_s$ and since the first term in the expansion 
of this quantity is of order 
$a_s^0$, we recover from (\ref{Wsol}) the well-known 
result that the LL solution for $C_9$ is induced only by mixing and 
is of order $1/a_s$.

\end{appendix}

\end{document}